\documentstyle[preprint,aps]{revtex}
\begin{document}
\tightenlines
\draft

\title
{Dirac Fields on Spacelike Hypersurfaces, Their Rest-Frame 
Description and Dirac Observables}

\author{Francesco Bigazzi}

\address
{Dipartimento di Fisica\\
Universita' di Milano\\
Via G.Celoria  16\\
20133 Milano, Italy}

\author{and}

\author{Luca Lusanna}

\address
{Sezione INFN di Firenze\\
L.go E.Fermi 2 (Arcetri)\\
50125 Firenze, Italy\\
E-mail LUSANNA@FI.INFN.IT}

\maketitle

\begin{abstract}

Grassmann-valued Dirac fields together with the electromagnetic field (the
pseudoclassical basis of QED) are reformulated on spacelike hypersurfaces in
Minkowski spacetime and then restricted to Wigner hyperplanes to get their
description in the rest-frame Wigner-covariant instant form of dynamics. The
canonical reduction to the Wigner-covariant Coulomb gauge is done in the
rest frame. It is shown, on the basis of a geometric incosistency, that the
description of fermions is incomplete, because there is no bosonic carrier of 
the spin structure describing the trajectory of the electric current in 
Minkowski spacetime, as it was already emphasized in connection with the first 
quantization of spinning particles in a previous paper.

\vskip 1truecm
\noindent July 1998
\vskip 1truecm

\end{abstract}
\vfill\eject

\section
{Introduction}

In a series of papers\cite{lusa,lv1,lv2,lv3} inspired by Ref.\cite{dira} the 
canonical reduction to a generalized noncovariant Coulomb gauge of the
standard SU(3)xSU(2)xU(1) model of elementary particles was obtained by using
the Shanmugadhasan canonical transformation \cite{sha} [see Refs.\cite{re} for
reviews].

In Ref.\cite{lus1,lus2} there was the definition of a new type of instant form 
of dynamics, the rest-frame 1-time Wigner-covariant instant form, which 
generalizes to special relativity the canonical separation of the center of mass
from the relative variables of an isolated system. It required the 
reformulation of classical isolated systems on arbitrary spacelike 
hypersurfaces foliating Minkowski spacetime (covariant 3+1 splitting), in a way
which is suited to the coupling to the gravitational field. The canonical 
reduction of tetrad gravity\cite{dplr} will, then, open the path to get a 
unified Hamiltonian description of the four interactions.

Therefore, one now has nearly all the technology needed to reduce the standard
model to a rest-frame Wigner-covariant generalized Coulomb gauge. This has been 
done for its bosonic part in Refs.\cite{lus1,lus2,lus3}, where the coupling of 
positive energy charged scalar particles plus the electromagnetic and color 
Yang-Mills fields was studied in the rest-frame Wigner-covariant instant form.

If one makes the canonical reduction of the gauge degrees of freedom of the
isolated system in the rest-frame instant form on the Wigner hyperplane, one 
gets the rest-frame Wigner-covariant generalized Coulomb gauge in which the
universal breaking of covariance is restricted to the decoupled center-of-mass
variable. However, as shown in Refs.\cite{lusa,lus1,lus3}, the region of 
spacetime over which this noncovariance is spread, is finite in spacelike 
directions and identifies a classical intrinsic unit of length, the M\o ller 
radius $\rho =\sqrt{-W^2}/P^2=|{\hat {\vec S}}| /\sqrt{P^2}$, where $P^2 > 0$ 
and $W^2=-P^2 {\hat {\vec S}}^2$ are the Poincar\'e Casimirs and ${\hat {\vec 
S}}$ the Thomas rest-frame spin of the isolated system respectively. This
unit of length gives rise to a physical intrinsic ultraviolet cutoff at the
quantum level in the spirit of Dirac and Yukawa.

To get the description of the standard model on spacelike hypersurfaces one
still needs the formulation of fermions on them. This is also needed for
treating the fermions in general relativity: given their coupling to tetrad
gravity (see for instance Refs.\cite{wei,naka}) one needs this formulation
to arrive at the ADM canonical formalism based on 3+1 splittings of the globally
hyperbolic asymptotically flat spacetime. However, as can be seen in
Refs.\cite{geh}, where there is known on the subject, in general one restricts
himself to hyperplanes $x^o=const.$ and there is no real discussion of the
Dirac brackets associated with the second class constraints.

As a first step, in a previous paper\cite{big} there was the study of positive
energy charged spinning particles plus the electromagnetic field, to which
we refer for a review of the approach and for the problematic concerning the
spin structure of these particles.

In this paper we shall study the formulation of Grassmann-valued Dirac fields
plus the electromagnetic field on spacelike hypersurfaces in Minkowski
spacetime. This is the pseudoclassical basis of QED.

In Section II there is the description of Grassmann-valued Dirac and Maxwell
fields on spacelike hypersurfaces in Minkowski spacetime.

In Section III there is their restriction to arbitrary spacelike hyperplanes,
while in Section IV there is the rest-frame description on Wigner
hyperplanes.

In SEction V there is the canonical reduction of the system to the
Wigner-covariant Coulomb gauge: a canonical basis of Dirac's observables is 
found and the reduced Hamilton equations are determined.

In the Conclusions there are some comments on the incompleteness of the
description of fermions and on their quantization.

In Appendix A there is a review on the foliations of Minkowski spacetime with
spacelike hypersurfaces.

In Appendix B there is a discussion on the Lagrangian for Dirac's fields on
spacelike hypersurfaces.

In Appendix C there are the transformation properties of Dirac's fields under 
Wigner boosts.

\vfill\eject

\section{Dirac and Maxwell Fields on Spacelike Hypersurfaces.}

As in Ref.\cite{lus1}, let us consider a 3+1 splitting of Minkowski spacetime 
with a family of spacelike hypersurfaces $\Sigma_{\tau}$, whose points $z^{\mu}
(\tau ,\vec \sigma )$ are labelled by Lorentz-scalar parameters: i) $\tau$ 
labelling the leaves of the foliation; ii) $\vec \sigma$ giving curvilinear 
coordinates on each leave. The coordinates $z^{\mu}(\tau ,\vec \sigma )$ will
be the fields describing the hypersurface. See Appendix A for a review of
notations and concepts, in particular for the flat tetrads $z^{\mu}_{\check A}
(\tau ,\vec \sigma )$ and their inverse cotetrads $z^{\check A}_{\mu}
(\tau ,\vec \sigma )$, connected to the 3+1 splittings.

On the hypersurface $\Sigma_{\tau}$, we describe the electromagnetic potential
and field strength with Lorentz-scalar variables $A_{\check A}(\tau ,\vec \sigma
)$ and $F_{{\check A}{\check B}}(\tau ,\vec \sigma )$ respectively, defined by

\begin{eqnarray}
&&A_{\check A}(\tau ,\vec \sigma )=z^{\mu}_{\check A}(\tau ,\vec \sigma )
A_{\mu}(z(\tau ,\vec \sigma )),\nonumber \\
&&F_{{\check A}{\check B}}(\tau ,\vec \sigma )={\partial}_{\check A}A_{\check
B}(\tau ,\vec \sigma )-{\partial}_{\check B}A_{\check A}(\tau ,\vec \sigma )=
z^{\mu}_{\check A}(\tau ,\vec \sigma )z^{\nu}_{\check B}(\tau ,\vec \sigma )
F_{\mu\nu}(z(\tau ,\vec \sigma )),
\label{II1}
\end{eqnarray}

\noindent and knowing the embedding of the spacelike hypersurfaces in Minkowski 
spacetime by construction.

The Grassmann-valued Dirac field on $\Sigma_{\tau}$ will be

\begin{equation}
\tilde \psi (\tau, \vec{\sigma})   \equiv \psi (z(\tau, \vec{\sigma})),
~~~~~~~~{\bar {\tilde \psi}} (\tau, \vec{\sigma})
\equiv \bar{\psi} (z(\tau, \vec{\sigma}))=\psi^{\dagger}(\tau ,\vec \sigma )
\gamma^o.
\label{II2}
\end{equation}

Since the field is Grassmann-valued,
its components $\psi_{\alpha}(\tau ,\vec \sigma )$, 
[$\alpha =1,..,4$ are spinor indices], satisfy

\begin{eqnarray}
{\tilde \psi}_{\alpha} (\tau, \vec{\sigma})  \bar{\tilde{\psi}}_{\beta} (\tau, 
\vec{\sigma}) +\bar{\tilde{\psi}}_{\beta}(\tau, \vec{\sigma}) {\tilde \psi}
_{\alpha} (\tau, \vec{\sigma}) = 0, \nonumber \\
{\tilde \psi}_{\alpha} (\tau, \vec{\sigma})  {\tilde \psi}_{\beta} (\tau, 
\vec{\sigma}) +{\tilde \psi}_{\beta} (\tau, \vec{\sigma}) {\tilde \psi}
_{\alpha} (\tau, \vec{\sigma}) = 0, \nonumber \\
\bar{\tilde{\psi}}_{\alpha} (\tau, \vec{\sigma})  \bar{\tilde{\psi}}_{\beta} 
(\tau, \vec{\sigma}) +\bar{\tilde{\psi}}_{\beta} (\tau, \vec{\sigma}) 
\bar{\tilde{\psi}}_{\alpha} (\tau, \vec{\sigma}) = 0 .
\label{II3}
\end{eqnarray}

In Appendix B there is a discussion of the coupling of tetrad gravity to 
fermion fields and of how to reexpress it after a 3+1 splitting of spacetime
so to be able to define the Hamiltonian formalism. After a restriction to
Minkowski spacetime, with the previous 3+1 splitting, we get the form of the
Lagrangian for Dirac fields on spacelike hypersurfaces $\Sigma_{\tau}$
[see Eq.(\ref{b3})]

\begin{eqnarray}
{\cal L}(\tau, \vec{\sigma})&=& N(\tau, \vec{\sigma})
 \sqrt{\gamma (\tau, \vec{\sigma})}
 \big[ \frac{i}{2}\bar{\tilde{\psi}} (\tau, \vec{\sigma}) \gamma^\mu
z^{\breve{A}}_\mu (\tau, \vec{\sigma})
\left( {\partial_{\breve{A}}} - i e A_{\breve{A}} (\tau, \vec{\sigma}) \right)
\tilde \psi (\tau, \vec{\sigma})  +  \nonumber \\
&-& \frac{i}{2} \left( {\partial_{\breve{A}}} + i e A_{\breve{A}}
(\tau, \vec{\sigma}) \right) \bar{\tilde{\psi}}(\tau, \vec{\sigma})
z^{\breve{A}}_\mu (\tau, \vec{\sigma}) \gamma^\mu \tilde \psi (\tau, 
\vec{\sigma})- m \bar{\tilde{\psi}} (\tau, \vec{\sigma})  \tilde \psi (\tau, 
\vec{\sigma}) \big] + \nonumber \\
&-& \frac{N(\tau, \vec{\sigma}) \sqrt{\gamma(\tau,
\vec{\sigma})}}{4} g^{\breve{A} \breve{C}} (\tau, \vec{\sigma})
g^{\breve{B} \breve{D}} (\tau, \vec{\sigma}) F_{\breve{A}
\breve{B}} (\tau, \vec{\sigma}) F_{\breve{C} \breve{D}} (\tau,
\vec{\sigma}).
\label{II4}
\end{eqnarray}

It is convenient to take as Lagrangian variables [since we will not use any more
the notation $\psi (z(\tau ,\vec \sigma ))$, no confusion will arise]

\begin{eqnarray}
\tilde \psi &\rightarrow & \psi = \, ^4{\! \! \! \! \! \sqrt{\gamma}}\, 
\tilde \psi , \nonumber \\
{\bar{\tilde \psi}}&\rightarrow &\bar{\psi} = \, ^4{\! \! \! \! \! 
\sqrt{\gamma}} \bar{\tilde{\psi}} .
\label{II5}
\end{eqnarray}

\noindent Since on spacelike hyperplanes\cite{lus1} one has $\gamma (\tau 
,\vec \sigma )=1$, there we shall recover $\psi =\tilde \psi$.

Eq.(\ref{II4}) becomes [see Appendix A for the definition of the lapse and
shift functions $N$, $N^{\check r}$; see also Ref.\cite{lus3}]

\begin{eqnarray}
{\cal L}(\tau, \vec{\sigma})&=& N(\tau, \vec{\sigma})
 \big[ \frac{i}{2}
\bar{\psi} (\tau, \vec{\sigma}) \gamma^\mu
z^{\breve{A}}_\mu (\tau, \vec{\sigma})
\left( {\partial_{\breve{A}}} - i e A_{\breve{A}} (\tau, \vec{\sigma}) \right)
\psi (\tau, \vec{\sigma})  +  \nonumber \\
&-& \frac{i}{2} \left( {\partial_{\breve{A}}} + i e A_{\breve{A}}
(\tau, \vec{\sigma}) \right)
\bar{\psi}(\tau, \vec{\sigma})
z^{\breve{A}}_\mu (\tau, \vec{\sigma}) \gamma^\mu \psi (\tau, \vec{\sigma})
- m \bar{\psi} (\tau, \vec{\sigma})  \psi (\tau, \vec{\sigma}) \big] + 
\nonumber \\
&-& \frac{N(\tau, \vec{\sigma}) \sqrt{\gamma(\tau,
\vec{\sigma})}}{4} g^{\breve{A} \breve{C}} (\tau, \vec{\sigma})
g^{\breve{B} \breve{D}} (\tau, \vec{\sigma}) F_{\breve{A}
\breve{B}} (\tau, \vec{\sigma}) F_{\breve{C} \breve{D}} (\tau,
\vec{\sigma})=\nonumber \\
&=&\int d\tau d^3\sigma [{i\over 2}l_{\mu} (\bar \psi \gamma^{\mu}(\partial
_{\tau}-ieA_{\tau})\psi -(\partial_{\tau}+ieA_{\tau})\bar \psi 
\gamma^{\mu}\psi)-\nonumber \\
&-&{i\over 2}N^{\check r}l_{\mu}(\bar \psi \gamma^{\mu}
(\partial_{\check r}-ieA_{\check r})\psi -(\partial_{\check r}+ieA_{\check r})
\bar \psi \gamma^{\mu}\psi )+\nonumber \\
&+&{i\over 2} N \gamma^{\check r\check s}z_{\check s\mu} (\bar \psi \gamma
^{\mu}(\partial_{\check r}-ieA_{\check r})\psi -\nonumber \\
&-&(\partial_{\check r}+ieA_{\check r})\bar \psi \gamma^{\mu}\psi )-
N m \bar \psi \psi ](\tau ,\vec \sigma )-\nonumber \\
&-&[{{\sqrt{\gamma}}\over {2N}}(F_{\tau \check r}-
N^{\check u}F_{\check u\check r})\gamma^{\check r\check s}(F_{\tau \check s}-
N^{\check v}F_{\check v\check s})+\nonumber \\
&+&{{N\sqrt{\gamma}}\over 4}\gamma^{\check r\check s}\gamma^{\check u\check v}
F_{\check r\check u}F_{\check s\check v}](\tau ,\vec \sigma ).
\label{II6}
\end{eqnarray}

The canonical momenta are [$E_{\check r}=F_{{\check r}\tau}$ and $B_{\check r}
={1\over 2}\epsilon_{{\check r}{\check s}{\check t}}F_{{\check s}{\check t}}$ 
($\epsilon_{{\check r}{\check s}{\check t}}=\epsilon^{{\check r}{\check s}
{\check t}}$) are the electric and magnetic fields respectively; for
$g_{\check A\check B}\rightarrow \eta_{\check A\check B}$ one gets 
$\pi^{\check r}=-E_{\check r}=E^{\check r}$]

\begin{eqnarray}
\pi_{\alpha}(\tau, \vec{\sigma}) &=& \frac{\partial {\cal L} (\tau, 
\vec{\sigma})}{\partial(\partial_\tau \psi_{\alpha})} =
- \frac i2 \left( \bar{\psi} (\tau, \vec{\sigma}) \gamma^\mu \right)_{\alpha}
l_\mu (\tau, \vec{\sigma}) ,\nonumber \\
\bar{\pi}_{\alpha} (\tau, \vec{\sigma}) &=&
\frac{\partial {\cal L} (\tau, \vec{\sigma})}
{\partial(\partial_\tau \bar{\psi}_{\alpha})} =
- \frac i2 \left( \gamma^\mu \psi (\tau, \vec{\sigma}) \right)^a
l_\mu (\tau, \vec{\sigma}), \nonumber \\
\pi^\tau (\tau, \vec{\sigma}) &=&
\frac{\partial^L {\cal L} (\tau, \vec{\sigma})}
{\partial(\partial_\tau A_\tau)} = 0, \nonumber \\
\pi^{\breve{r}} (\tau, \vec{\sigma}) &=&
 \frac{\partial^L {\cal L} (\tau, \vec{\sigma})}
{\partial(\partial_\tau A_{\breve{r}})} =
- \frac{\gamma (\tau, \vec{\sigma})}{\sqrt{g(\tau, \vec{\sigma})}}
\gamma^{\breve{r} \breve{s}} (\tau, \vec{\sigma})
\big( F_{\tau \breve{s}} - g_{\tau \breve{v}}
\gamma^{\breve{v} \breve{u}} F_{\breve{u} \breve{s}}\big)
(\tau, \vec{\sigma})= \nonumber \\
&=&{ {\gamma (\tau ,\vec \sigma )}\over {\sqrt {g(\tau ,\vec \sigma )}} }
\gamma^{{\check r}{\check s}}(\tau ,\vec \sigma )(E_{\check s}(\tau ,\vec 
\sigma )+g_{\tau {\check v}}(\tau ,\vec \sigma )\gamma^{{\check v}{\check u}}
(\tau ,\vec \sigma )\epsilon_{{\check u}{\check s}{\check t}} B_{\check t}
(\tau ,\vec \sigma )),\nonumber \\
\rho_{\mu} (\tau, \vec{\sigma}) &=&
 - \frac{\partial^L {\cal L} (\tau, \vec{\sigma})}
{\partial z^\mu_\tau} = l_\mu (\tau, \vec{\sigma})
\Big\{ - \frac i2 \gamma^{\breve{r} \breve{s}} (\tau, \vec{\sigma})
z_{\nu \breve{s}} (\tau, \vec{\sigma})\Big[
\bar{\psi} (\tau, \vec{\sigma}) \gamma^\nu \partial_{\breve{r}}
\psi (\tau, \vec{\sigma}) + \nonumber \\
&-& \partial_{\breve{r}} \bar{\psi} (\tau, \vec{\sigma}) \gamma^\nu
\psi (\tau, \vec{\sigma})\Big] + m \bar{\psi} (\tau, \vec{\sigma})
\psi (\tau, \vec{\sigma}) + \nonumber \\
 &-& \frac{1}{2\sqrt{\gamma (\tau, \vec{\sigma})}}
\pi^{\breve{r}} (\tau, \vec{\sigma})
g_{\breve{r} \breve{s}}(\tau, \vec{\sigma})
\pi^{\breve{s}} (\tau, \vec{\sigma})+ \nonumber \\
&+& \frac {\sqrt{\gamma (\tau, \vec{\sigma})}}{4}
\gamma^{\breve{r} \breve{s}} (\tau, \vec{\sigma})
\gamma^{\breve{u} \breve{v}} (\tau, \vec{\sigma})
F_{\breve{r} \breve{u}} (\tau, \vec{\sigma})
F_{\breve{s} \breve{v}} (\tau, \vec{\sigma}) + \nonumber \\
&-& e \gamma^{\breve{r} \breve{s}} (\tau, \vec{\sigma})
z_{\nu \breve{s}} (\tau, \vec{\sigma})
 A_{\breve{r}} (\tau, \vec{\sigma})
\bar{\psi} (\tau, \vec{\sigma}) \gamma^\nu
\psi (\tau, \vec{\sigma}) \Big\} + \nonumber \\
&+& z_{\mu \breve{s}} (\tau, \vec{\sigma})
\gamma^{\breve{r} \breve{s}} (\tau, \vec{\sigma})
\Big\{ \frac i2 l_\nu (\tau, \vec{\sigma})
\Big[
\bar{\psi} (\tau, \vec{\sigma}) \gamma^\nu \partial_{\breve{r}}
\psi (\tau, \vec{\sigma}) + \nonumber \\
&-& \partial_{\breve{r}} \bar{\psi} (\tau, \vec{\sigma}) \gamma^\nu
\psi (\tau, \vec{\sigma})\Big] + \nonumber \\
&+& F_{\breve{r} \breve{u}} (\tau, \vec{\sigma})
\pi^{\breve{u}} (\tau, \vec{\sigma}) +
e  A_{\breve{r}} (\tau, \vec{\sigma})
\bar{\psi} (\tau, \vec{\sigma}) \gamma^\nu l_\nu (\tau, \vec{\sigma})
\psi (\tau, \vec{\sigma}) \Big\}.
\label{II7}
\end{eqnarray}

They satisfy the Poisson brackets

\begin{eqnarray}
\{ \psi_{\alpha} (\tau , \vec{\sigma}), \pi_{\beta} (\tau , 
\vec{\sigma}^\prime) \} &=& \{ \pi_{\beta} (\tau , \vec{\sigma}^\prime) , 
\psi_{\alpha} (\tau , \vec{\sigma}) \} = - \delta_{\alpha\beta} \delta^3 
(\vec{\sigma} - \vec{\sigma}^\prime) ,\nonumber \\
\{ \bar{\psi}_{\alpha} (\tau , \vec{\sigma}),
\bar{\pi}_{\beta} (\tau , \vec{\sigma}^\prime) \}
&=& \{ \bar{\pi}_{\beta} (\tau , \vec{\sigma}^\prime) ,
\bar{\psi}_{\alpha} (\tau , \vec{\sigma}) \}
= - \delta_{\alpha\beta} \delta^3 (\vec{\sigma} - \vec{\sigma}^\prime) ,
\nonumber \\
\{ z^\mu (\tau , \vec{\sigma}), \rho_\nu (\tau , \vec{\sigma}^\prime) \}
&=& - \eta^\mu_\nu \delta^3 (\vec{\sigma} - \vec{\sigma}^\prime) ,\nonumber \\
\{ A_{\breve{A}} (\tau , \vec{\sigma}),
\pi^{\breve{B}} (\tau , \vec{\sigma}^\prime) \}
&=& \eta^{\breve{B}}_{\breve{A}}
 \delta^3 (\vec{\sigma} - \vec{\sigma}^\prime) .
\label{II8}
\end{eqnarray}

The primary constraints are

\begin{eqnarray}
\chi_{\alpha} (\tau , \vec{\sigma}) &\equiv &
\pi_{\alpha} (\tau , \vec{\sigma}) + \frac i2 \left(\bar{\psi} (\tau , 
\vec{\sigma})\gamma^\mu \right)_{\alpha} l_\mu (\tau , \vec{\sigma}) 
\approx 0 ,\nonumber \\
\bar{\chi}_{\alpha} (\tau , \vec{\sigma}) &\equiv &
\bar{\pi}_{\alpha} (\tau , \vec{\sigma}) + \frac i2
\left( \gamma^\mu \psi (\tau , \vec{\sigma})
\right)_{\alpha} l_\mu (\tau , \vec{\sigma}) \approx 0, \nonumber \\
\pi^\tau (\tau , \vec{\sigma}) &\approx & 0 ,\nonumber \\
{\cal H}_\mu (\tau , \vec{\sigma}) &\equiv &
\rho_\mu (\tau , \vec{\sigma}) -
l_\mu (\tau, \vec{\sigma})
\Big\{ - \frac i2 \gamma^{\breve{r} \breve{s}} (\tau, \vec{\sigma})
z_{\nu \breve{s}} (\tau, \vec{\sigma})\Big[
\bar{\psi} (\tau, \vec{\sigma}) \gamma^\nu \partial_{\breve{r}}
\psi (\tau, \vec{\sigma}) + \nonumber \\
&-& \partial_{\breve{r}} \bar{\psi} (\tau, \vec{\sigma}) \gamma^\nu
\psi (\tau, \vec{\sigma})\Big] + m \bar{\psi} (\tau, \vec{\sigma})
\psi (\tau, \vec{\sigma}) + \nonumber \\
 &-& \frac{1}{2\sqrt{\gamma (\tau, \vec{\sigma})}}
\pi^{\breve{r}} (\tau, \vec{\sigma})
g_{\breve{r} \breve{s}}(\tau, \vec{\sigma})
\pi^{\breve{s}} (\tau, \vec{\sigma})+ \nonumber \\
&+& \frac {\sqrt{\gamma (\tau, \vec{\sigma})}}{4}
\gamma^{\breve{r} \breve{s}} (\tau, \vec{\sigma})
\gamma^{\breve{u} \breve{v}} (\tau, \vec{\sigma})
F_{\breve{r} \breve{u}} (\tau, \vec{\sigma})
F_{\breve{s} \breve{v}} (\tau, \vec{\sigma}) + \nonumber \\
&-& e \gamma^{\breve{r} \breve{s}} (\tau, \vec{\sigma})
z_{\nu \breve{s}} (\tau, \vec{\sigma})
 A_{\breve{r}} (\tau, \vec{\sigma})
\bar{\psi} (\tau, \vec{\sigma}) \gamma^\nu
\psi (\tau, \vec{\sigma}) \Big\} + \nonumber \\
&+& \gamma^{\breve{r} \breve{s}} (\tau, \vec{\sigma})
z_{\mu \breve{s}} (\tau, \vec{\sigma})
\Big\{ \frac i2 l_\nu (\tau, \vec{\sigma})
\Big[
\bar{\psi} (\tau, \vec{\sigma}) \gamma^\nu \partial_{\breve{r}}
\psi (\tau, \vec{\sigma}) + \nonumber \\
&-& \partial_{\breve{r}} \bar{\psi} (\tau, \vec{\sigma}) \gamma^\nu
\psi (\tau, \vec{\sigma})\Big]+ \nonumber \\
&+& F_{\breve{r} \breve{u}} (\tau, \vec{\sigma})
\pi^{\breve{u}} (\tau, \vec{\sigma}) +
e  A_{\breve{r}} (\tau, \vec{\sigma})
\bar{\psi} (\tau, \vec{\sigma}) \gamma^\nu l_\nu (\tau, \vec{\sigma})
\psi (\tau, \vec{\sigma}) \Big\} \approx 0 .
\label{II9}
\end{eqnarray}

\noindent In ${\cal H}_{\mu}(\tau, \vec{\sigma})\approx 0$ the coefficient of
$l_{\mu}(\tau, \vec{\sigma})$ is the energy density $T^{\tau\tau}(\tau, 
\vec{\sigma})$ of the isolated system, while the coefficient of $z_{\check r\mu}
(\tau, \vec{\sigma})$ is the 3-momentum density $T^{\tau \check r}(\tau, 
\vec{\sigma})$.

The canonical and Dirac Hamiltonians are

\begin{eqnarray}
H_c &=& \int{d^3\sigma \Big[ - \pi_{\alpha} (\tau , \vec{\sigma})
\partial_\tau \psi_{\alpha} (\tau , \vec{\sigma}) -
\bar{\pi}_{\alpha} (\tau , \vec{\sigma})
\partial_\tau \bar{\psi}_{\alpha} (\tau , \vec{\sigma}) +
\pi^{\breve{A}} (\tau , \vec{\sigma})
\partial_\tau A_{\breve{A}} (\tau , \vec{\sigma}) }+ \nonumber \\
 & - & \rho_\mu (\tau , \vec{\sigma})
z^\mu_\tau (\tau , \vec{\sigma}) - {\cal L} (\tau , \vec{\sigma}) \Big] =
- \int{d^3\sigma \Gamma (\tau , \vec{\sigma})
A_\tau (\tau , \vec{\sigma})} ,\nonumber \\
H_D &=& \int{d^3\sigma \Big[ - A_\tau (\tau , \vec{\sigma})
\Gamma (\tau , \vec{\sigma}) + \lambda^\mu (\tau , \vec{\sigma})
{\cal H}_\mu (\tau , \vec{\sigma}) +
{\bar a}_{\alpha} (\tau , \vec{\sigma}) \chi_{\alpha} (\tau , \vec{\sigma})}+ 
\nonumber \\
&+&  \bar{\chi}_{\alpha} (\tau , \vec{\sigma})a_{\alpha} (\tau , \vec{\sigma}) +
\mu_\tau (\tau , \vec{\sigma}) \pi^\tau (\tau , \vec{\sigma})\Big] ,
\label{II10}
\end{eqnarray}

\noindent where $\lambda^\mu$, $\mu_\tau$, $a_{\alpha}$ are Dirac multipliers
[$a_{\alpha}$ are odd multipliers]. 

The constraints $\chi_{\alpha}$, ${\bar \chi}_{\alpha}$, are second class,
because they satisfy the Poisson brackets

\begin{equation}
\{ \chi_{\alpha} (\tau , \vec{\sigma}) ,
\bar{\chi}_{\beta} (\tau , \vec{\sigma}^\prime) \} =
-i \left( \gamma^\mu l_\mu (\tau , \vec{\sigma})\right)_{\beta\alpha}
\delta^3 (\vec{\sigma} - \vec{\sigma}^\prime) .
\label{II11}
\end{equation}

\noindent It is not convenient to go to Dirac brackets with respect to them at 
this stage, because the fundamental variables would not have any more diagonal 
Dirac brackets; the elimination of these constraints will be delayed till when 
the theory will be restricted to spacelike hyperplanes.

The constraints ${\cal H}_\mu (\tau , \vec{\sigma}) \approx 0$ imply that the
description is independent from the choice of the family of spacelike 
hypersurfaces used to foliate Minkowski spacetime. Since these constraints
do not have zero Poisson brackets with $\chi_{\alpha}$, ${\bar \chi}_{\alpha}$

\begin{eqnarray}
\{ {\cal H}_{\perp} (\tau , \vec{\sigma}),
\chi_{\alpha} (\tau , \vec{\sigma}^\prime) \} &=&
i \gamma^{\breve{r} \breve{s}} (\tau , \vec{\sigma})
z_{\mu \breve{s}} (\tau , \vec{\sigma})
\big( \partial_{\breve{r}} \bar{\psi} (\tau , \vec{\sigma}) \gamma^\mu \big)
_{\alpha}\delta^3 (\vec{\sigma} - \vec{\sigma}^\prime) + \nonumber \\
&+& \frac i2 \big( \bar{\psi} (\tau , \vec{\sigma}) \gamma^\mu \big)_{\alpha}
\partial_{\breve{r}} \big( \gamma^{\breve{r} \breve{s}}
z_{\mu \breve{s}}\big) (\tau , \vec{\sigma})
\delta^3 (\vec{\sigma} - \vec{\sigma}^\prime) + \nonumber \\
&+& m \bar{\psi}_{\alpha} (\tau , \vec{\sigma})
\delta^3 (\vec{\sigma} - \vec{\sigma}^\prime) + \nonumber \\
&-& e \gamma^{\breve{r} \breve{s}} (\tau , \vec{\sigma})
z_{\mu \breve{s}} (\tau , \vec{\sigma}) A_{\breve{r}} (\tau , \vec{\sigma})
\big( \bar{\psi} (\tau , \vec{\sigma}) \gamma^\mu \big)_{\alpha}
\delta^3 (\vec{\sigma} - \vec{\sigma}^\prime) , \nonumber \\
\{ {\cal H}_{\perp} (\tau , \vec{\sigma}),
\bar{\chi}_{\alpha} (\tau , \vec{\sigma}^\prime) \} &=&
i \gamma^{\breve{r} \breve{s}} (\tau , \vec{\sigma})
z_{\mu \breve{s}} (\tau , \vec{\sigma})
\big( \gamma^\mu \partial_{\breve{r}}
{\psi} (\tau , \vec{\sigma}) \big)_{\alpha}
\delta^3 (\vec{\sigma} - \vec{\sigma}^\prime) + \nonumber \\
&+& \frac i2 \big( \gamma^\mu \psi (\tau , \vec{\sigma}) \big)_{\alpha}
\partial_{\breve{r}} \big( \gamma^{\breve{r} \breve{s}}
z_{\mu \breve{s}}\big) (\tau , \vec{\sigma})
\delta^3 (\vec{\sigma} - \vec{\sigma}^\prime) + \nonumber \\
&-& m \psi_{\alpha} (\tau , \vec{\sigma})
\delta^3 (\vec{\sigma} - \vec{\sigma}^\prime) + \nonumber \\
&+& e \gamma^{\breve{r} \breve{s}} (\tau , \vec{\sigma})
z_{\mu \breve{s}} (\tau , \vec{\sigma}) A_{\breve{r}} (\tau , \vec{\sigma})
\big( \gamma^\mu \psi (\tau , \vec{\sigma}) \big)_{\alpha}
\delta^3 (\vec{\sigma} - \vec{\sigma}^\prime) , \nonumber \\
\{ {\cal H}_{\breve{r}} (\tau , \vec{\sigma}),
\chi_{\alpha} (\tau , \vec{\sigma}^\prime) \} &=&
- i \partial_{\breve{r}}
\big(\bar{\psi} (\tau , \vec{\sigma}) \gamma^\mu \big)_{\alpha}
l_\mu (\tau , \vec{\sigma})
\delta^3 (\vec{\sigma} - \vec{\sigma}^\prime) + \nonumber \\
&+& \frac i2 \big( \bar{\psi} (\tau , \vec{\sigma}^\prime) \gamma^\mu \big)
_{\alpha}l_{\mu } (\tau , \vec{\sigma}^\prime) \partial_{\breve{r}}
\delta^3 (\vec{\sigma} - \vec{\sigma}^\prime) +  \nonumber \\
&+& e A_{\breve{r}} (\tau , \vec{\sigma})
\big( \bar{\psi} (\tau , \vec{\sigma}) \gamma^\mu \big)_{\alpha}
l_\mu (\tau , \vec{\sigma})
\delta^3 (\vec{\sigma} - \vec{\sigma}^\prime) ,\nonumber \\
\{ {\cal H}_{\breve{r}} (\tau , \vec{\sigma}),
\bar{\chi}_{\alpha} (\tau , \vec{\sigma}^\prime) \} &=&
- i \partial_{\breve{r}}
\big( \gamma^\mu \psi (\tau , \vec{\sigma}) \big)_{\alpha}
l_\mu (\tau , \vec{\sigma})
\delta^3 (\vec{\sigma} - \vec{\sigma}^\prime) + \nonumber \\
&+& \frac i2 \big( \gamma^\mu \psi (\tau , \vec{\sigma}^\prime) \big)_{\alpha}
l_{\mu } (\tau , \vec{\sigma}^\prime) \partial_{\breve{r}}
\delta^3 (\vec{\sigma} - \vec{\sigma}^\prime) +  \nonumber \\
&-& e A_{\breve{r}} (\tau , \vec{\sigma})
\big( \gamma^\mu \psi (\tau , \vec{\sigma}) \big)_{\alpha}
l_\mu (\tau , \vec{\sigma})
\delta^3 (\vec{\sigma} - \vec{\sigma}^\prime) ,
\label{II12}
\end{eqnarray}

\noindent where

\begin{eqnarray}
{\cal H}_{\perp} (\tau , \vec{\sigma})
&\equiv & l^\mu (\tau , \vec{\sigma})
{\cal H}_\mu (\tau , \vec{\sigma}) \approx 0, \nonumber \\
{\cal H}_{\breve{r}} (\tau , \vec{\sigma})
 &\equiv & z_{\breve{r}}^\mu (\tau , \vec{\sigma})
{\cal H}_\mu (\tau , \vec{\sigma}) \approx 0,
\label{II13}
\end{eqnarray}

\noindent it is convenient to introduce the new constraints

\begin{eqnarray}
{\cal H}_\mu^\ast (\tau , \vec{\sigma}) &=&
{\cal H}_\mu (\tau , \vec{\sigma}) -
\int{d^3u \{ {\cal H}_{\mu} (\tau , \vec{\sigma}),
\bar{\chi}_{\beta} (\tau , \vec{u}) \}
i \big( \gamma^\mu l_\mu (\tau , \vec{u})\big)_{\alpha\beta}
\chi_{\alpha} (\tau, \vec{u})} + \nonumber \\
&-& \int{ d^3u \{ {\cal H}_{\mu} (\tau , \vec{\sigma}),
{\chi}_{\beta} (\tau , \vec{u}) \}
i \big( \gamma^\mu l_\mu (\tau , \vec{u})\big)_{\beta\alpha}
\bar{\chi}_{\alpha} (\tau, \vec{u})} \approx {\cal H}_\mu (\tau , \vec{\sigma})
\approx 0,\nonumber \\
&&{}\nonumber \\
\{ {\cal H}_\mu^\ast (\tau , \vec{\sigma}) ,
\chi_{\alpha} (\tau , \vec{\sigma}^\prime) \} &\approx& 0, ~~~~~
\{ {\cal H}_\mu^\ast (\tau , \vec{\sigma}) ,
\bar{\chi}_{\alpha} (\tau , \vec{\sigma}^\prime) \} \approx 0 ,\nonumber \\
\{ {\cal H}_\mu^\ast (\tau , \vec{\sigma}) ,
{\cal H}_\nu^\ast (\tau , \vec{\sigma}^\prime) \} &\approx &
\Biggl\{ \biggl[ l_\mu (\tau , \vec{\sigma})
z_{\breve{r} \nu} (\tau , \vec{\sigma}) -
l_\nu (\tau , \vec{\sigma})
z_{\breve{r} \mu} (\tau , \vec{\sigma}) \biggr]
\frac{\pi^{\breve{r}} (\tau , \vec{\sigma})}
{\sqrt{\gamma (\tau , \vec{\sigma})}} + \nonumber \\
&-& z_{\breve{u} \mu} (\tau , \vec{\sigma})
\gamma^{\breve{u} \breve{r}} (\tau , \vec{\sigma})
F_{\breve{r} \breve{s}} (\tau , \vec{\sigma})
\gamma^{\breve{s} \breve{v}} (\tau , \vec{\sigma})
z_{\breve{v} \nu} (\tau , \vec{\sigma}) \Biggr\} \cdot \nonumber \\
&\cdot & \Gamma (\tau , \vec{\sigma})
\delta^3 (\vec{\sigma} - \vec{\sigma}^\prime) \approx 0.
\label{II14}
\end{eqnarray}

\noindent where $\Gamma (\tau ,\vec \sigma )\approx 0$, the Gauss law
constraint, will be defined in the next equation.
Therefore the constraints ${\cal H}_{\mu}^{*}(\tau ,\vec \sigma )\approx 0$
are constants of the motion and first class.

The time constancy of $\pi^{\tau}(\tau ,\vec \sigma )\approx 0$, i.e.
$\partial_{\tau}\, \pi^{\tau}(\tau ,\vec \sigma )\, {\buildrel \circ \over =}\,
\{ \pi^{\tau}(\tau ,\vec \sigma ),H_D\} \approx 0$ [${\buildrel \circ \over
=}$ means evaluated on the solution of the equations of motion], gives the Gauss
law secondary constraint

\begin{equation}
\Gamma (\tau ,\vec \sigma )=\partial_{\check r}\, \pi^{\check r}(\tau ,\vec 
\sigma )+e\bar \psi (\tau ,\vec \sigma )\gamma^{\mu}l_{\mu}(\tau ,\vec \sigma )
\psi (\tau ,\vec \sigma )\approx 0.
\label{II15}
\end{equation}

\noindent Since we have

\begin{eqnarray}
&&\{ \Gamma (\tau ,\vec \sigma ),\chi_{\alpha}(\tau ,{\vec \sigma}^{'}) \} =-
e {\bar \psi}_{\beta}(\tau ,\vec \sigma ) (\gamma^{\mu}l_{\mu}(\tau ,\vec 
\sigma ) )_{\beta\alpha} \delta^3(\vec \sigma -{\vec \sigma}^{'}),
\nonumber \\
&&\{ \Gamma (\tau ,\vec \sigma ),{\bar \chi}_{\alpha}(\tau ,{\vec \sigma}^{'})
\} = e (\gamma^{\mu}l_{\mu}(\tau ,\vec \sigma ) )_{\alpha\beta} \psi_{\beta}
(\tau ,\vec \sigma ) \delta^3(\vec \sigma -{\vec \sigma}^{'}),
\label{II16}
\end{eqnarray}

\noindent let us define 

\begin{equation}
\Gamma^{*}(\tau ,\vec \sigma )\equiv \Gamma (\tau ,\vec \sigma )+ie[\chi
_{\alpha} \psi_{\alpha}+{\bar \psi}_{\alpha} {\bar \chi}_{\alpha}]
(\tau ,\vec \sigma )\approx 0.
\label{II17}
\end{equation}

\noindent This new constraint satisfies

\begin{eqnarray}
&&\{ \Gamma^{*}(\tau ,\vec \sigma ),\chi_{\alpha}(\tau ,{\vec \sigma}^{'})\}
=-ie \chi_{\alpha}(\tau ,\vec \sigma )\delta^3(\vec \sigma -{\vec \sigma}^{'}),
\nonumber \\
&&\{ \Gamma^{*}(\tau ,\vec \sigma ),{\bar \chi}_{\alpha}(\tau ,{\vec \sigma}
^{'}) \} =ie {\bar \chi}_{\alpha}(\tau ,\vec \sigma )\delta^3(\vec \sigma -
{\vec \sigma}^{'}),\nonumber \\
&&\{ \Gamma^{*}(\tau ,\vec \sigma ),{\cal H}^{*}_{\mu}(\tau ,{\vec \sigma}^{'})
\} \approx 0,\nonumber \\
&&\{ \Gamma^{*}(\tau ,\vec \sigma ),\Gamma^{*}(\tau ,{\vec \sigma}^{'}) \}
\approx 0,\nonumber \\
&&\{ \Gamma^{*}(\tau ,\vec \sigma ),\pi^{\tau}(\tau ,{\vec \sigma}^{'}) \}
=0.
\label{II18}
\end{eqnarray}

\noindent Therefore, $\Gamma^{*}(\tau ,\vec \sigma )\approx 0$ is a constant of
motion and a first class constraint like $\pi^{\tau}(\tau ,\vec \sigma )
\approx 0$.

The time constancy of $\chi_{\alpha}$, ${\bar \chi}_{\alpha}$

\begin{eqnarray}
\partial_\tau \chi_{\alpha} (\tau , \vec{\sigma})\, &{\buildrel \circ \over
=}&\, \{ \chi_{\alpha} (\tau , 
\vec{\sigma}),H_D\} = -e A_\tau (\tau , \vec{\sigma})\, 
\big( \bar{\psi} (\tau , \vec{\sigma}) \gamma^\mu \big)_{\alpha}
l_\mu (\tau , \vec{\sigma}) + \nonumber \\
&+& i {\bar a}_{\beta} (\tau , \vec{\sigma}) \big( \gamma^\mu
l_\mu (\tau , \vec{\sigma})\big)_{\beta \alpha}\approx 0, \nonumber \\
\partial_\tau \bar{\chi}_{\alpha} (\tau , \vec{\sigma})\,
&{\buildrel \circ \over =}&\, 
\{ \bar{\chi}_{\alpha} (\tau , \vec{\sigma}), H_D\} =
 e A_\tau (\tau , \vec{\sigma})\,
l_\mu (\tau , \vec{\sigma})
\big( \gamma^\mu \psi (\tau , \vec{\sigma}) \big)_{\alpha}- \nonumber \\
&-& i a_{\beta} (\tau , \vec{\sigma}) \big( \gamma^\mu
l_\mu (\tau , \vec{\sigma})\big)_{\alpha\beta}\approx 0,
\label{II19}
\end{eqnarray}

\noindent gives

\begin{equation}
{\bar a}_{\alpha} (\tau , \vec{\sigma}) \chi_{\alpha} (\tau , \vec{\sigma}) +
a_{\alpha} (\tau , \vec{\sigma}) \bar{\chi}_{\alpha} (\tau , \vec{\sigma}) 
\approx -i e A_\tau (\tau , \vec{\sigma})
\big( \chi_{\alpha} \psi_{\alpha} +\bar{\psi}_{\alpha} \bar{\chi}_{\alpha} 
\big) (\tau , \vec{\sigma}),
\label{II20}
\end{equation}

\noindent so that the final Dirac Hamiltonian is

\begin{equation}
H_D = \int{d^3\sigma \Big[ -A_\tau (\tau , \vec{\sigma})
\Gamma^\ast (\tau , \vec{\sigma}) + \lambda^\mu (\tau , \vec{\sigma})
{\cal H}^\ast_\mu (\tau , \vec{\sigma}) +
\mu_\tau (\tau , \vec{\sigma}) \pi^\tau (\tau , \vec{\sigma}) \Big]}
\label{II21}
\end{equation}

\noindent in which only the first class constraints ${\cal H}^{*}_{\mu}$, 
$\pi^{\tau}$, $\Gamma^{*}$ appear.

One can show that the 10 Poincar\'e generators

\begin{eqnarray}
{p_s}_\mu &=& \int{d^3\sigma \rho_\mu (\tau , \vec{\sigma})}, \nonumber \\
J^{\mu \nu} &=& \int{d^3\sigma
\Big[ z^\mu (\tau , \vec{\sigma}) \rho^\nu (\tau , \vec{\sigma})
- z^\nu (\tau , \vec{\sigma}) \rho^\mu (\tau , \vec{\sigma})\Big]} + 
\nonumber \\
&+& \frac{i}{2} \int{d^3\sigma
\Big[ \pi (\tau , \vec{\sigma}) \sigma^{\mu \nu} \psi (\tau , \vec{\sigma})
+ \bar{\psi} (\tau , \vec{\sigma}) \sigma^{\mu \nu}
\bar{\pi} (\tau , \vec{\sigma})\Big]},
\label{II22}
\end{eqnarray}

\noindent and the electric charge [see Ref.\cite{lusa} for the boundary 
conditions on the fields and for the extraction of this weak charge from 
the Gauss law first class constraint]

\begin{equation}
Q = - e \int{d^3\sigma \Big[ \bar{\psi} (\tau , \vec{\sigma})
\gamma^\mu l_\mu (\tau , \vec{\sigma}) \psi (\tau , \vec{\sigma}) \Big] } ,
\label{II23}
\end{equation}

\noindent are constants of the motion.

Let us note that like $\gamma^o$ the matrix $\gamma^\mu l_\mu (\tau , 
\vec{\sigma})$  satisfies
$\big[ \gamma^\mu l_\mu (\tau , \vec{\sigma}) \big]^2 = {I}$ and that we
have $\{ z^\mu (\tau , \vec{\sigma}),
{p_s}^\nu \} = - \eta^{\mu \nu}$.

\vfill\eject

\section{Restriction to Spacelike Hyperplanes.}

As in Ref.\cite{lus1}, let us restrict ourselves to spacelike hyperplanes 
$\Sigma_{\tau H}$ by adding the gauge fixings

\begin{eqnarray}
& \zeta^\mu (\tau , \vec{\sigma}) = z^\mu (\tau , \vec{\sigma})
- x_s^\mu (\tau) - b^\mu_{\breve{r}} (\tau) \sigma^{\breve{r}}
\approx 0 , \nonumber \\
& \{ \zeta^\mu (\tau , \vec{\sigma}),
{\cal H}^\ast_\nu (\tau , \vec{\sigma}^\prime) \} =
- \eta^\mu_\nu \delta^3 (\vec{\sigma} - \vec{\sigma}^\prime) ,
\label{III1}
\end{eqnarray}

\noindent and the Dirac brackets

\begin{eqnarray}
\{ A, B \}^\ast &=& \{ A, B \} -
\int{d^3\sigma \Big[ \{ A, \zeta^\mu (\tau , \vec{\sigma}) \}
\{ {\cal H}^\ast_\mu (\tau , \vec{\sigma}) , B \} }+ \nonumber \\
&-& \{ A, {\cal H}^\ast_\mu (\tau , \vec{\sigma}) \}
\{ \zeta^\mu (\tau , \vec{\sigma}), B \} \Big] .
\label{III2}
\end{eqnarray}

The hyperplane $\Sigma_{\tau H}$ is described by 10 configuration variables: 
an origin $x^{\mu}_s(\tau )$ and the 6 independent degrees of freedom in an 
orthonormal tetrad $b^{\mu}_{\check A}(\tau )$ [${}b^{\mu}_{\check A}\, \eta
_{\mu\nu} b^{\nu}_{\check B}=\eta_{\check A\check B}$] with $b^{\mu}_{\tau}=l
^{\mu}$, where $l^{\mu}$ is the $\tau$-independent normal to the hyperplane.
Now, we have $z^{\mu}_{\check r}(\tau ,\vec \sigma )\equiv b^{\mu}_{\check r}
(\tau )$, $z^{\mu}_{\tau}(\tau ,\vec \sigma )\equiv {\dot x}^{\mu}_s(\tau )+
b^{\mu}_{\check r}(\tau ) \sigma^{\check r}$, $g_{\check r\check s}(\tau ,\vec 
\sigma )\equiv -\delta_{\check r\check s}$, $\gamma^{\check r\check s}
(\tau ,\vec \sigma )\equiv -\delta^{\check r\check s}$, $\gamma (\tau ,\vec 
\sigma )=det\, g_{\check r\check s}(\tau ,\vec \sigma )\equiv 1$. The
nonvanishing Dirac brackets of the variables $x^\mu_s$,
$p_s^\mu$, $b^\mu_{\breve{A}}$, $S^{\mu \nu}_s$, $A_{\breve{A}}$,
$\pi^{\breve{A}}$,  are 
[$C^{\mu\nu\alpha\beta}_{\gamma\delta}=\eta^{\nu}_{\gamma}\eta^{\alpha}
_{\delta}\eta^{\mu\beta}+\eta^{\mu}_{\gamma}\eta^{\beta}_{\delta}\eta
^{\nu\alpha}-\eta^{\nu}_{\gamma}\eta^{\beta}_{\delta}\eta^{\mu\alpha}-
\eta^{\mu}_{\gamma}\eta^{\alpha}_{\delta}\eta^{\nu\beta}$ are the structure 
constants of the Lorentz group]

\begin{eqnarray}
\{ x_s^\mu (\tau), p_s^\nu \}^\ast &=& -\eta^{\mu \nu}, \nonumber \\
\{ S^{\mu \nu}_s(\tau) , b^\rho_{\breve{A}}\}^\ast &=&
\eta^{\rho \nu} b^\mu_{\breve{A}} (\tau) -
\eta^{\rho \mu} b^\nu_{\breve{A}} (\tau), \nonumber \\
\{ S^{\mu \nu}_s(\tau) , S^{\alpha \beta}_s(\tau) \}^\ast &=&
C^{\mu \nu \alpha \beta}_{\gamma \delta} S^{\gamma \delta}_s (\tau).
\label{III3}
\end{eqnarray}

While $p^{\mu}_s$ is the momentum conjugate to $x^{\mu}_s$, the 6 independent
momenta conjugate to the 6 degrees of freedom in the $b^{\mu}_{\check A}$'s
are hidden in $S_s^{\mu\nu}$, which is a component of the angular momentum 
tensor

\begin{eqnarray}
J^{\mu \nu} &=& L^{\mu \nu}_s + S^{\mu \nu}_s + S^{\mu \nu}_{\psi} , 
\nonumber \\
L^{\mu \nu}_s &=& x^\mu_s(\tau) p^\nu_s - x^\nu_s(\tau) p^\mu_s , 
\nonumber \\
S^{\mu \nu}_s &=& b^\mu_{\breve{r}} (\tau) \int{d^3\sigma \sigma^{\breve{r}}
\rho^\nu (\tau, \vec{\sigma})} -
b^\nu_{\breve{r}} (\tau) \int{d^3\sigma \sigma^{\breve{r}}
\rho^\mu (\tau, \vec{\sigma})} , \nonumber \\
S^{\mu \nu}_{\psi}  &=& {i\over 2} \int d^3\sigma 
\Big[ \pi (\tau , \vec{\sigma}) \sigma^{\mu \nu}
\psi (\tau , \vec{\sigma}) +
\bar{\psi} (\tau , \vec{\sigma}) \sigma^{\mu \nu}
\bar{\pi} (\tau , \vec{\sigma}) \Big] ,\nonumber \\
&&{}\nonumber \\
&&\{ J^{\mu\nu},J^{\alpha\beta} \}^{*} = C^{\mu\nu\alpha\beta}_{\gamma\delta} 
J^{\gamma\delta},\quad\quad \{ L_s^{\mu\nu},L_s^{\alpha\beta} \}^{*} = C
^{\mu\nu\alpha\beta}_{\gamma\delta} L_s^{\gamma\delta},\nonumber \\
&&\{ S_s^{\mu\nu},S_s^{\alpha\beta} \}^{*} 
= C^{\mu\nu\alpha\beta}_{\gamma\delta} 
S_s^{\gamma\delta},\quad\quad \{ S_{\xi}^{\mu\nu},S_{\xi}^{\alpha\beta} \}^{*}
= C^{\mu\nu\alpha\beta}_{\gamma\delta} S_{\xi}^{\gamma\delta}.
\label{III4}
\end{eqnarray}

The relations

\begin{eqnarray}
\{ S^{\mu \nu}_\psi, \psi_{\alpha} (\tau , \vec{\sigma}) \}^\ast &=&
\frac i2 \big( \sigma^{\mu \nu} \psi (\tau , \vec{\sigma}) \big)
_{\alpha}, \nonumber \\
\{ S^{\mu \nu}_\psi, \bar{\psi}_{\alpha} (\tau , \vec{\sigma}) \}^\ast &=&
- \frac i2 \big(\bar{\psi} (\tau , \vec{\sigma}) \sigma^{\mu \nu} \big)
_{\alpha},
\label{III5}
\end{eqnarray}

\noindent show that $S^{\mu\nu}_{\psi}$ is the generator of the symplectic 
action of the Lorentz transformations on the Dirac fields.

Since by asking the time constancy of the gauge fixings (\ref{III1}) 
we get\cite{lus1} $\lambda^{\mu}(\tau ,\vec \sigma )={\tilde \lambda}
^{\mu}(\tau )+{\tilde \lambda}^{\mu}{}_{\nu}(\tau )b^{\nu}_{\check r}(\tau ) 
\sigma^{\check r}$, ${\tilde \lambda}^{\mu}(\tau )=-{\dot x}^{\mu}_s(\tau )$, 
${\tilde \lambda}^{\mu\nu}(\tau )=-{\tilde \lambda}^{\nu\mu}(\tau )={1\over 2}
\sum_{\check r}[{\dot b}^{\mu}_{\check r}b^{\nu}_{\check r}-b^{\mu}_{\check r}
{\dot b}^{\nu}_{\check r}](\tau )$, the Dirac Hamiltonian becomes

\begin{equation}
H^F_D = \tilde{\lambda}^\mu(\tau) \tilde{H}_\mu^{*}(\tau) -
\frac 12 \tilde{\lambda}^{\mu \nu}(\tau) \tilde{H}^{*}_{\mu \nu}(\tau) +
\int{d^3\sigma \Big[ - A_\tau (\tau, \vec{\sigma}) \Gamma^{*}(\tau, 
\vec{\sigma}) +
\mu_\tau (\tau, \vec{\sigma}) \pi^\tau (\tau, \vec{\sigma}) \Big] } ,
\label{III6}
\end{equation}

\noindent where

\begin{eqnarray}
\tilde{H}^\ast_\mu (\tau) &=& \int{d^3\sigma
{\cal H}^\ast_\mu (\tau , \vec{\sigma})} \approx 0 ,\nonumber \\
\tilde{H}^\ast_{\mu \nu}(\tau) &=&
b_{\mu \breve{r}} (\tau) \int{d^3\sigma
\sigma^{\breve{r}} {\cal H}^\ast_\nu (\tau , \vec{\sigma})} -
b_{\nu \breve{r}} (\tau) \int{d^3\sigma
\sigma^{\breve{r}} {\cal H}^\ast_\mu (\tau , \vec{\sigma})} \approx 0.
\label{III7}
\end{eqnarray}

On $\Sigma_{\tau H}$ the second class constraints have the form

\begin{eqnarray}
&\chi_{\alpha} (\tau , \vec{\sigma}) = \pi_{\alpha} (\tau , \vec{\sigma}) +
\frac i2 \big( \bar{\psi} (\tau , \vec{\sigma})
 \gamma^\mu \big)_{\alpha} b_{\mu \tau} \approx 0 & ,\nonumber \\
&\bar{\chi}_{\alpha} (\tau , \vec{\sigma}) = \bar{\pi}_{\alpha} (\tau , 
\vec{\sigma}) +\frac i2 \big( \gamma^\mu {\psi} (\tau , \vec{\sigma})
 \big)_{\alpha} b_{\mu \tau} \approx 0 & ,\nonumber \\
&\{ \chi_{\alpha} (\tau , \vec{\sigma}),
\bar{\chi}_{\beta} (\tau , \vec{\sigma}^\prime)\}^\ast =
-i \big( \gamma^\mu b_{\mu \tau} \big)_{\beta\alpha}
\delta^3 (\vec{\sigma} - \vec{\sigma}^\prime)& ,\nonumber \\
&\Big[\big( \gamma^\mu b_{\mu \tau} \big)^2 = {I} \Big] .  &
\label{III8}
\end{eqnarray}

\noindent and can be eliminated by introducing the Dirac brackets

\begin{eqnarray}
\{ A, B \}^\ast_D &=& \{ A, B \}^\ast -
\int{d^3u \{ A, \chi_{\alpha} (\tau , \vec{u}) \}^\ast i
\big( \gamma^\mu b_{\mu \tau} \big)_{\alpha\beta}
\{ \bar{\chi}_{\beta} (\tau , \vec{u}), B \}^\ast }+ \nonumber \\
&-& \int{d^3u \{ A, \bar{\chi}_{\alpha} (\tau , \vec{u}) \}^\ast i
\big( \gamma^\mu b_{\mu \tau} \big)_{\beta\alpha}
\{ {\chi}_{\beta} (\tau , \vec{u}), B \}^\ast } .
\label{III9}
\end{eqnarray}

Now we get ${\tilde H}^{*}_{\mu}(\tau )\equiv H_{\mu}(\tau )=\int d^3\sigma
{\cal H}_{\mu}(\tau, \vec{\sigma})$, ${\tilde H}_{\mu\nu}^{*}(\tau )\equiv 
{\tilde H}_{\mu\nu}(\tau )=b_{\mu \breve{r}} (\tau) \int{d^3\sigma
\sigma^{\breve{r}} {\cal H}^\ast_\nu (\tau , \vec{\sigma})} -
b_{\nu \breve{r}} (\tau) \int{d^3\sigma
\sigma^{\breve{r}} {\cal H}^\ast_\mu (\tau , \vec{\sigma})}$,

\begin{equation}
S^{\mu \nu}_\psi \equiv
\frac 14 \int{d^3\sigma b_{\rho \tau} \bar{\psi}(\tau , \vec{\sigma})
\big[ \gamma^\rho , \sigma^{\mu \nu} \big]_+ \psi (\tau , \vec{\sigma})},
\label{III10}
\end{equation}

\noindent and

\begin{equation}
\{ \psi_{\alpha} (\tau , \vec{\sigma}) ,
\bar{\psi}_{\beta} (\tau , \vec{\sigma}^\prime) \}^\ast_D =
- i \big( \gamma^\mu b_{\mu \tau} \big)_{\alpha\beta}
\delta^3 (\vec{\sigma} - \vec{\sigma}^\prime) ,
\label{III11}
\end{equation}

\noindent while the Dirac brackets of the variables $x^\mu_s$,
$p_s^\mu$, $b^\mu_{\breve{A}}$, $A_{\breve{A}}$, $\pi^{\breve{A}}$ are left
unaltered by the new brackets. Now only the total spin 

\begin{equation}
S^{\mu \nu} = S^{\mu \nu}_s + S^{\mu \nu}_\psi ,
\label{III12}
\end{equation}

\noindent satisfies a Lorentz algebra, since we have

\begin{eqnarray}
\{ S^{\mu \nu} , S^{\alpha \beta} \}^\ast_D &=&
C^{\mu \nu \alpha \beta}_{\gamma \delta} S^{\gamma \delta}, \nonumber \\
\{ S^{\mu \nu} , \psi_{\alpha} (\tau , \vec{\sigma}) \}^\ast_D &=&
\frac i2 \big( \sigma^{\mu \nu} \psi (\tau , \vec{\sigma}) \big)_{\alpha}, 
\nonumber \\
\{ S^{\mu \nu} , \bar{\psi}_{\alpha} (\tau , \vec{\sigma}) \}^\ast_D &=&
- \frac i2 \big( \bar{\psi} (\tau , \vec{\sigma})
\sigma^{\mu \nu} \big)_{\alpha},\nonumber \\
&&{}\nonumber \\
\{ L^{\mu \nu}_s , L^{\alpha \beta}_s \}^\ast_D &=&
\{ L^{\mu \nu}_s , L^{\alpha \beta}_s \}^\ast =
C^{\mu \nu \alpha \beta}_{\gamma \delta} L_s^{\gamma \delta}, \nonumber \\
\{ L^{\mu \nu}_s , S^{\alpha \beta} \}^\ast_D &=&
\{ L^{\mu \nu}_s , S^{\alpha \beta} \}^\ast = 0,\nonumber \\
\{ J^{\mu \nu} , J^{\alpha \beta} \}^\ast_D &=&
\{ J^{\mu \nu} , J^{\alpha \beta} \}^\ast =
C^{\mu \nu \alpha \beta}_{\gamma \delta} J^{\gamma \delta} .
\label{III13}
\end{eqnarray}

The Dirac Hamiltonian becomes

\begin{eqnarray}
H_D &=& \tilde{\lambda}^\mu (\tau) \tilde{H}_\mu (\tau) -
\frac 12 \tilde{\lambda}^{\mu \nu} (\tau)
\tilde{H}_{\mu \nu}(\tau) + \nonumber \\
&+& \int{d^3\sigma \Big[ - A_\tau (\tau , \vec{\sigma})
\Gamma (\tau , \vec{\sigma})  + \mu_\tau (\tau , \vec{\sigma})
\pi^\tau (\tau , \vec{\sigma}) \Big]} ,
\label{III14}
\end{eqnarray}

\noindent and contains all the first class constraints

\begin{eqnarray}
\Gamma (\tau , \vec{\sigma}) &=&
\partial_{\breve{r}} \pi^{\breve{r}} (\tau , \vec{\sigma}) +
e \bar{\psi} (\tau , \vec{\sigma}) \gamma^\mu b_{\mu \tau}
\psi (\tau , \vec{\sigma}) \approx 0, \nonumber \\
\pi^\tau (\tau , \vec{\sigma}) &\approx & 0, \nonumber \\
\tilde{H}_\mu (\tau) &=&
 \int{d^3\sigma {\cal H}_\mu (\tau , \vec{\sigma})} = {p_s}_\mu
- b_{\mu \tau} \int{d^3\sigma \Big[ \frac i2
b_{\nu \breve{r}} (\tau) \Big(
\bar{\psi} (\tau , \vec{\sigma}) \gamma^\nu \partial_{\breve{r}}
\psi (\tau , \vec{\sigma})} + \nonumber \\
&-& \partial_{\breve{r}} \bar{\psi} (\tau , \vec{\sigma}) \gamma^\nu
\psi (\tau , \vec{\sigma}) \Big) +
m \bar{\psi} (\tau , \vec{\sigma}) \psi (\tau , \vec{\sigma}) +
\frac{\vec{\pi}^2 (\tau , \vec{\sigma}) + \vec{B}^2 (\tau , \vec{\sigma})}
{2} + \nonumber \\
&+& e b_{\nu \breve{r}} (\tau) A_{\breve{r}} (\tau , \vec{\sigma})
\bar{\psi} (\tau , \vec{\sigma})
\gamma^\nu \psi (\tau , \vec{\sigma}) \Big] + \nonumber \\
&+& b_{\mu \breve{r}} (\tau)
\int{d^3\sigma \Big[ \frac {i}{2}
b_{\nu \tau} (\tau) \Big(
\bar{\psi} (\tau , \vec{\sigma}) \gamma^\nu \partial_{\breve{r}}
\psi (\tau , \vec{\sigma})} + \nonumber \\
 &-&
 \partial_{\breve{r}} \bar{\psi} (\tau , \vec{\sigma}) \gamma^\nu
\psi (\tau , \vec{\sigma}) \Big) +
 \Big( \vec{\pi} (\tau , \vec{\sigma}) \wedge
 \vec{B} (\tau , \vec{\sigma}) \Big)_{\breve{r}} + \nonumber \\
 &+&
 e A_{\breve{r}} (\tau , \vec{\sigma})
\bar{\psi} (\tau , \vec{\sigma})
\gamma^\nu b_{\nu \tau} \psi (\tau , \vec{\sigma}) \Big] \approx 0, \nonumber \\
\tilde{H}^{\mu \nu} (\tau) &=&
\int{d^3\sigma b_{\breve{r}}^\mu (\tau) \sigma^{\breve{r}}
{\cal H}^\nu (\tau , \vec{\sigma})} -
\int{d^3\sigma b_{\breve{r}}^\nu (\tau) \sigma^{\breve{r}}
{\cal H}^\mu (\tau , \vec{\sigma})} = \nonumber \\
&=& S^{\mu \nu} (\tau) - S^{\mu \nu}_\psi (\tau) + \nonumber \\
&-& \Big(
b_{\breve{r}}^\mu (\tau) b^\nu_\tau - b_{\breve{r}}^\nu (\tau) b^\mu_\tau \Big)
\int{d^3\sigma \sigma^{\breve{r}} \Big[ \frac i2
b_{\nu \breve{s}} (\tau) \Big( \bar{\psi} (\tau , \vec{\sigma})
\gamma^\nu \partial_{\breve{s}}
\psi (\tau , \vec{\sigma})} + \nonumber \\
&-& \partial_{\breve{s}} \bar{\psi} (\tau , \vec{\sigma}) \gamma^\nu
\psi (\tau , \vec{\sigma}) \Big) +
 m \bar{\psi} (\tau , \vec{\sigma}) \psi (\tau , \vec{\sigma}) +
\frac{\vec{\pi}^2 (\tau , \vec{\sigma}) + \vec{B}^2 (\tau , \vec{\sigma})}
{2} + \nonumber \\
&+& e b_{\nu \breve{s}} (\tau) A_{\breve{s}} (\tau , \vec{\sigma})
\bar{\psi} (\tau , \vec{\sigma})
\gamma^\nu \psi (\tau , \vec{\sigma}) \Big] + \nonumber \\
&+& \Big( b_{\breve{r}}^\mu (\tau) b_{\breve{s}}^\nu (\tau) -
b_{\breve{r}}^\nu (\tau) b_{\breve{s}}^\mu (\tau) \Big)
\int{d^3\sigma \sigma^{\breve{r}} \Big[ \frac i2
b_{\rho \tau} \Big( \bar{\psi} (\tau , \vec{\sigma}) \gamma^\rho
 \partial_{\breve{s}} \psi (\tau , \vec{\sigma})} + \nonumber \\
 &-& \partial_{\breve{s}} \bar{\psi} (\tau , \vec{\sigma}) \gamma^\rho
\psi (\tau , \vec{\sigma}) \Big) +
 \Big( \vec{\pi} (\tau , \vec{\sigma}) \wedge
 \vec{B} (\tau , \vec{\sigma}) \Big)_{\breve{s}} + \nonumber \\
 &+&  e A_{\breve{s}} (\tau , \vec{\sigma})
\bar{\psi} (\tau , \vec{\sigma})
\gamma^\rho b_{\rho \tau} \psi (\tau , \vec{\sigma}) \Big] \approx 0,
\label{III15}
\end{eqnarray}

\noindent with [see Eq.(\ref{III14})]

\begin{equation}
\{ {\cal H}^\ast_\mu (\tau , \vec{\sigma}),
{\cal H}^\ast_\nu (\tau , \vec{\sigma}^\prime) \} =
\{ {\cal H}^\ast_\mu (\tau , \vec{\sigma}),
{\cal H}^\ast_\nu (\tau , \vec{\sigma}^\prime) \}^\ast
\approx \{ {\cal H}_\mu (\tau , \vec{\sigma}),
{\cal H}_\nu (\tau , \vec{\sigma}^\prime) \}^\ast_D\approx 0 .
\label{III16}
\end{equation}

\vfill\eject

\section{The Rest-Frame Descrption on Wigner's Hyperplanes.}

The next step \cite{lus1} is to select all the configurations of the isolated 
system which are timelike, namely with $p^2_s > 0$. For them we can boost at 
rest with the standard Wigner boost $L^\mu_{.\nu}(\stackrel{o}p_s, p_s)$ for
timelike Poincar\'e orbits all the variables of the noncanonical basis
$x_s^\mu(\tau)$, $p_s^\mu$ , $b_{\breve{A}}^\mu(\tau )$ , $S_s^{\mu \nu}(\tau 
)$ , $A_{\breve{A}}(\tau ,\vec \sigma )$ , $\pi^{\breve{A}}(\tau ,\vec \sigma 
)$, $\psi (\tau ,\vec \sigma )$, $\bar \psi (\tau ,\vec \sigma )$
 with Lorentz indices (except $p^{\mu}_s$).

Let us now introduce the new variables [see Ref.\cite{lus1} and Appendix C;
they are obtained with a canonical transformation $e^{\{ .,{\cal F}(p_s)\} }$,
whose generator is given in Eq.(\ref{c3})]

\begin{eqnarray}
b^A_{\breve{B}} (\tau) &=&
\epsilon^A_\mu ( u(p_s)) b^\mu_{\breve{B}} (\tau) =
L^A_{.\mu} (\stackrel{o}p_s, p_s) b^\mu_{\breve{B}} (\tau), \nonumber \\
\tilde{x}^\mu_s (\tau) &=& x^\mu_s (\tau) - \frac 12
\epsilon^A_\nu ( u(p_s)) \eta_{AB}
\frac{\partial \epsilon^B_\rho (u(p_s))} {\partial {p_s}_\mu}
S^{\nu \rho} =  \nonumber \\
&=& x^\mu_s (\tau) - \frac {1} {\eta_s \sqrt{p^2_s}
(p_s^0 + \eta_s \sqrt{p^2_s})}
\Big[ {p_s}_\nu S^{\nu \mu} + \eta_s \sqrt{p^2_s}
\Big( S^{0 \mu} - S^{0 \nu} \frac{ {p_s}_\nu p_s^\mu} {p^2_s} \Big) \Big] = 
\nonumber \\
&=& x^\mu_s (\tau) - \frac {1} {\eta_s \sqrt{p^2_s}}
\Big[ \eta^\mu_A \Big( \bar{S}^{\bar{o} A} -
\frac{ \bar{S}^{Ar} p^r_s} {(p_s^0 + \eta_s \sqrt{p^2_s})} \Big) + \nonumber \\
&+& \frac{p_s^\mu + 2 \eta_s \sqrt{p^2_s} \eta^{\mu 0}}
{\eta_s \sqrt{p^2_s}
(p_s^0 + \eta_s \sqrt{p^2_s})} \bar{S}^{\bar{o} r}_s p^r_s \Big] ,\nonumber \\
p^\mu_s &=& p^\mu_s,~~~~~A_{\breve{A}} (\tau, \vec{\sigma}) =
A_{\breve{A}} (\tau, \vec{\sigma}), ~~~~~ \pi^{\breve{A}} (\tau, \vec{\sigma})
= \pi^{\breve{A}} (\tau, \vec{\sigma}) ,\nonumber \\
\tilde{S}^{\mu \nu} &=& S^{\mu \nu} +
\frac 12 \epsilon^A_\rho ( u(p_s)) \eta_{AB}
\Big( \frac{\partial \epsilon^B_\sigma (u(p_s))} {\partial {p_s}_\mu} p^\nu_s
- \frac{\partial \epsilon^B_\sigma (u(p_s))} {\partial {p_s}_\nu} p^\mu_s \Big)
S^{\rho \sigma} = \nonumber \\
&=&  S^{\mu \nu} + \frac {1} {\eta_s \sqrt{p^2_s}
(p_s^0 + \eta_s \sqrt{p^2_s})} \Big[ p_{s \beta} (S^{\beta \mu} p_s^\nu -
S^{\beta \nu} p_s^\mu ) + \eta_s \sqrt{p^2_s}
(S^{0 \mu} p_s^\nu - S^{0 \nu} p_s^\mu )\Big] ,\nonumber \\
\stackrel{o} \psi (\tau, \vec{\sigma}) &=& S( L(\stackrel{o}p_s, p_s))
\psi (\tau, \vec{\sigma}), \nonumber \\
\stackrel{o} {\bar{\psi}} (\tau, \vec{\sigma}) &=&
\bar{\psi} (\tau, \vec{\sigma}) S^{-1}( L(\stackrel{o}p_s, p_s)),
\label{IV1}
\end{eqnarray}

\noindent where $\eta_s = sign~p^0_s$ and

\begin{equation}
\bar{S}^{AB} = \epsilon^A_\mu ( u(p_s)) \epsilon^A_\nu ( u(p_s))S^{\mu \nu},
\label{IV2}
\end{equation}

\noindent  the ``rest-frame spin tensor" (or ``Thomas spin tensor") 
[ the $\bar{S}^{AB} $'s satisfy a Lorentz algebra]. Now the Lorentz generators
are

\begin{equation}
J^{\mu \nu} = \tilde{L}^{\mu \nu} + \tilde{S}^{\mu \nu} =
\tilde{x}^\mu_s p^\nu_s - \tilde{x}^\nu_s p^\mu_s + \tilde{S}^{\mu \nu} .
\label{IV3}
\end{equation}

The new variables have the following Dirac brackets [see Appendix C]

\begin{eqnarray}
\{ \tilde{x}^\mu_s , p^\nu_s \}^\ast_D &=& - \eta^{\mu \nu}, \nonumber \\
\{ \tilde{x}^\mu_s , \stackrel{o} \psi (\tau, \vec{\sigma}) \}^\ast_D &=&
\{ \tilde{x}^\mu_s , \stackrel{o}{\bar{\psi}} (\tau, \vec{\sigma}) \}^\ast_D =
\{ \tilde{x}^\mu_s , \tilde{x}^\nu_s \}^\ast_D =
\{ \tilde{x}^\mu_s , b^A_{\breve{B}} \}^\ast_D = 0 ,\nonumber \\
\{ \tilde{S}^{0i} , b^r_{\breve{A}} \}^\ast_D &=&
\frac{\delta^{is} (p^r_s b^s_{\breve{A}} - p^s_s b^r_{\breve{A}})}
{(p_s^0 + \eta_s \sqrt{p^2_s})}, \nonumber \\
\{ \tilde{S}^{ij} , b^r_{\breve{A}} \}^\ast_D &=&
(\delta^{ir} \delta^{js} - \delta^{is} \delta^{jr} )b^s_{\breve{A}}, 
\nonumber \\
\{ \tilde{S}^{\mu \nu} , \tilde{S}^{\alpha \beta} \}^\ast_D &=&
C^{\mu \nu \alpha \beta}_{\gamma \delta} \tilde{S}^{\gamma \delta}, \nonumber \\
\{ \tilde{x}^\mu_s , \tilde{S}^{0i} \}^\ast_D &=&
- \frac{1}{(p_s^0 + \eta_s \sqrt{p^2_s})} \Big[ \eta^{\mu j} \tilde{S}^{ji} +
\frac{ (p_s^\mu + \eta^{\mu 0} \eta_s \sqrt{p^2_s} ) \tilde{S}^{ik} p^k}
{ \eta_s \sqrt{p^2_s}(p_s^0 + \eta_s \sqrt{p^2_s})} \Big] ,\nonumber \\
\{ \tilde{x}^\mu_s , \tilde{S}^{ij} \}^\ast_D &=& 0 ,\nonumber \\
\{ \stackrel{o} {\psi_{\alpha}} (\tau, \vec{\sigma}) ,
\stackrel{o}{\bar{\psi}_{\beta}} (\tau, \vec{\sigma}^\prime) \}^\ast_D &=&
- i b^A_\tau (\tau) \eta_{A \sigma} (\gamma^\sigma)_{\alpha\beta}
\delta^3 (\vec{\sigma} - \vec{\sigma}^\prime) , \nonumber \\
\{ \tilde{S}^{\mu \nu} , \stackrel{o} \psi (\tau, \vec{\sigma}) \}^\ast_D &=&
\frac i2 \Big[ L^\mu_{.\alpha}(p_s, \stackrel{o}p_s)
L^\nu_{.\beta}(p_s, \stackrel{o}p_s)
- \frac 12 \epsilon^A_\rho ( u(p_s)) \eta_{AB}
 \Big( \frac{\partial \epsilon^B_\sigma (u(p_s))}
{\partial {p_s}_\mu} p^\nu_s + \nonumber \\
&-& \frac{\partial \epsilon^B_\sigma (u(p_s))} {\partial {p_s}_\nu} p^\mu_s 
\Big)
 L^\rho_{.\alpha}(p_s, \stackrel{o}p_s)
L^\sigma_{.\beta}(p_s, \stackrel{o}p_s) \Big] \sigma^{\alpha \beta}
\stackrel{o} \psi (\tau, \vec{\sigma}), \nonumber \\
\{ \tilde{S}^{\mu \nu} ,
\stackrel{o}{\bar{\psi}} (\tau, \vec{\sigma}) \}^\ast_D &=&
- \frac i2 \stackrel{o}{\bar{\psi}} (\tau, \vec{\sigma})
\Big[ L^\mu_{.\alpha}(p_s, \stackrel{o}p_s)
L^\nu_{.\beta}(p_s, \stackrel{o}p_s)
- \frac 12 \epsilon^A_\rho ( u(p_s)) \eta_{AB} \cdot \nonumber \\
&\cdot & \Big( \frac{\partial \epsilon^B_\sigma (u(p_s))}
{\partial {p_s}_\mu} p^\nu_s
- \frac{\partial \epsilon^B_\sigma (u(p_s))} {\partial {p_s}_\nu} p^\mu_s \Big)
 L^\rho_{.\alpha}(p_s, \stackrel{o}p_s)
L^\sigma_{.\beta}(p_s, \stackrel{o}p_s) \Big] \sigma^{\alpha \beta} .
\label{IV4}
\end{eqnarray}

Moreover, we have

\begin{equation}
\{ \bar{S}^{AB} ,\bar{S}^{CD}  \}^\ast_D =
C^{ABCD}_{EF} \bar{S}^{EF},
\label{IV5}
\end{equation}

\begin{eqnarray}
\{ \bar{S}^{AB} , \stackrel{o} \psi (\tau, \vec{\sigma}) \}^\ast_D &=&
\frac i2 \delta^A_{\alpha} \delta^B_{\beta} \sigma^{\alpha \beta}
\stackrel{o} \psi (\tau, \vec{\sigma}), \nonumber \\
\{ \bar{S}^{AB} , \stackrel{o}{\bar{\psi}} (\tau, \vec{\sigma}) \}^\ast_D &=&
- \frac i2 \stackrel{o}{\bar{\psi}} (\tau, \vec{\sigma})
\delta^A_{\alpha} \delta^B_{\beta} \sigma^{\alpha \beta} .
\label{IV6}
\end{eqnarray}

The new canonical origin $\tilde{x}^\mu_s (\tau)$ is not covariant, since under 
a Poincar\'e transformation $(a, \Lambda)$ it transforms as\cite{lus1}

\begin{equation}
\tilde{x}^\mu_s \stackrel{(a, \Lambda)} \longrightarrow 
\tilde{x}^{\prime \mu}_s =
\Lambda^\mu_{.\nu} \Big[ \tilde{x}^\nu_s +
\frac 12 \bar{S}_{rs} R^r_{.k} (\Lambda, p_s)
\frac{\partial}{\partial {p_s}_\nu} R^s_{.k} (\Lambda, p) \Big] + a^\mu .
\label{IV7}
\end{equation}

As shown in Ref.\cite{lus1}, we can restrict ourselves to the Wigner
hyperplane $\Sigma_{\tau W}$ with $l^{\mu}=u^{\mu}(p_s)$ [i.e. orthogonal to 
$p^{\mu}_s$] with the gauge fixings

\begin{eqnarray}
& T^\mu_{\breve{A}} (\tau) =
b^\mu_{\breve{A}} (\tau) - \epsilon^\mu_{A=\breve{A}} (u(p_s)) \approx 0 
\nonumber \\
& \Rightarrow b^A_{\breve{A}} (\tau) =
\epsilon_\mu^A (u(p_s)) b^\mu_{\breve{A}} (\tau) \approx \eta^A_{\breve{A}}.
\label{IV8}
\end{eqnarray}

\noindent whose time constancy implies ${\tilde \lambda}^{\mu\nu}(\tau )
\approx 0$. After having introduced the Dirac brackets

\begin{eqnarray}
\{ A , B \}^{\ast \ast}_D &=& \{ A , B \}^{\ast}_D
- \frac 14 \Big[ \{ A , \tilde{H}^{\gamma \delta} \}^{\ast}_D
\Big( \eta_{\gamma \sigma} \epsilon_\delta^D (u(p_s)) -
\eta_{\delta \sigma} \epsilon_\gamma^D (u(p_s))\Big)
\{ T^\sigma_D , B \}^{\ast}_D + \nonumber \\
&+& \{ A , T^\sigma_B\}^{\ast}_D
\Big( \eta_{\sigma \nu} \epsilon_\mu^B (u(p_s)) -
\eta_{\sigma \mu} \epsilon_\nu^B (u(p_s))\Big)
\{ \tilde{H}^{\mu \nu} , B \}^{\ast}_D \Big] ,
\label{IV9}
\end{eqnarray}

\noindent we get $b^{\mu}_{\check A}(\tau )\equiv
L^\mu{}_A(p_s,\stackrel{o}p_s)$ and ${\tilde H}^{\mu\nu}(\tau )\equiv 0$,
namely the determination of $S^{\mu\nu}_s=S^{\mu\nu}-S^{\mu\nu}_{\psi}$ in 
terms of the variables of the system. The remaining variables form a canonical 
basis

\begin{eqnarray}
\{ \tilde{x}^\mu_s(\tau), p_s^\nu \}^{\ast \ast}_D &=& - \eta^{\mu \nu}, 
\nonumber \\
\{ A_A(\tau ,\vec \sigma ),\pi^B(\tau ,{\vec \sigma}^{'}) \}^{\ast \ast}_D&=&
\eta^B_A \delta^3(\vec \sigma -{\vec \sigma}^{'}),\nonumber \\
\{ \stackrel{o}{\psi_{\alpha}} (\tau, \vec{\sigma}),
\stackrel{o}{\bar{\psi}}_{\beta} (\tau, \vec{\sigma}^\prime) \}^{\ast \ast}_D 
&=&-i (\gamma^{\bar{o}})_{\alpha\beta}
\delta^3 (\vec{\sigma} - \vec{\sigma}^\prime) .
\label{IV10}
\end{eqnarray}

As shown in Ref.\cite{lus1}, the dependence of the gauge-fixing (\ref{IV20}) on
$p^{\mu}_s$ implies that the Lorentz-scalar indices $\check A$ become Wigner
indices A: i) $A_{A=\tau}(\tau ,\vec \sigma )$ is a Lorentz-scalar field; ii)
$A_{A=r}(\tau ,\vec \sigma )$, 
are Wigner spin 1 ${}{}$ 3-vectors which transform with Wigner 
rotations under the action of Minkowski Lorentz boosts. In particular,
$\gamma^{\bar o}=\gamma^{\tau}$ is a Lorentz scalar matrix, while $\gamma^r$
form a Wigner spin 1 ${}$ 3-vector. Therefore, we have a Wigner-covariant 
realization of Dirac matrices

\begin{eqnarray}
&\gamma^A = \Big( \gamma^{\bar{o}}; \{ \gamma^r \} , r=1,2,3 \Big) ,\nonumber \\
&\big[ \gamma^A, \gamma^B \big]_+ = 2 \eta^{AB} ,
\label{IV11}
\end{eqnarray}

\noindent like in the Chakrabarti representation\cite{cha} (see also Ref.
\cite{big}). Under a Lorentz transformation $\Lambda$, the bilinears in the 
Dirac field transform with the associated Wigner rotation $R(\Lambda ,p_s)$.
For instance we have

\begin{eqnarray}
\stackrel{o}{\bar{\psi}} (\tau, \vec{\sigma}) \gamma^A
\stackrel{o}{\psi} (\tau, \vec{\sigma})&& \stackrel{\Lambda}\longrightarrow
R^A_{.B} (\Lambda, p_s) \stackrel{o}{\bar{\psi}} (\tau, \vec{\sigma}) \gamma^B
\stackrel{o}{\psi} (\tau, \vec{\sigma}),\nonumber \\
&&{}\nonumber \\
\stackrel{o}{\bar{\psi}} (\tau, \vec{\sigma}) \gamma^{\bar{o}}
\stackrel{o}{\psi} (\tau, \vec{\sigma}) &\stackrel{\Lambda}\longrightarrow &
\stackrel{o}{\bar{\psi}} (\tau, \vec{\sigma}) \gamma^{\bar{o}}
\stackrel{o}{\psi} (\tau, \vec{\sigma})~~~~~(scalar), \nonumber \\
\stackrel{o}{\bar{\psi}} (\tau, \vec{\sigma}) \gamma^r
\stackrel{o}{\psi} (\tau, \vec{\sigma}) &\stackrel{\Lambda}\longrightarrow  &
R^r_{.s} (\Lambda, p_s) \stackrel{o}{\bar{\psi}} (\tau, \vec{\sigma}) \gamma^s
\stackrel{o}{\psi} (\tau, \vec{\sigma})~~~(Wigner\, 3-vector),
\label{IV12}
\end{eqnarray}

\noindent and the induced spinorial transformation on Dirac fields will be
[$S(\Lambda ) \mapsto {\buildrel \circ \over S}(R(\Lambda ,p_s))$, which
could be evaluated by using the last of the next formulas]

\begin{eqnarray}
\stackrel{o}{\psi} (\tau, \vec{\sigma}) &\stackrel{\Lambda}\longrightarrow &
\stackrel{o}{\psi^\prime} (\tau, \vec{\sigma}) ={\buildrel \circ \over
S}(R(\Lambda, p_s))\stackrel{o}{\psi} (\tau, \vec{\sigma}) ,\nonumber \\
\stackrel{o}{\bar{\psi}} (\tau, \vec{\sigma}) &\stackrel{\Lambda}
\longrightarrow &
\stackrel{o}{\bar{\psi}^\prime} (\tau, \vec{\sigma}) =
\stackrel{o}{\bar{\psi}} (\tau, \vec{\sigma}) {\buildrel \circ \over S}^{-1}
(R(\Lambda, p_s)),\nonumber \\
&&{}\nonumber \\
{\buildrel \circ \over S}^{-1}(R(\Lambda, p_s)) &\gamma^A& {\buildrel \circ
\over S}(R(\Lambda, p_s))
= R^A_{.B} (\Lambda, p_s) \gamma^B.
\label{IV13}
\end{eqnarray}

The original variables $z^{\mu}(\tau, \vec{\sigma})$, $\rho_{\mu}(\tau, 
\vec{\sigma})$, are reduced only to ${\tilde x}^{\mu}_s(\tau )$, $p^{\mu}_s$,
on the Wigner hyperplane $\Sigma_{\tau W}$. On it there remain only six first 
class constraints

\begin{eqnarray}
\pi^\tau (\tau, \vec{\sigma}) &\approx & 0, \nonumber \\
\Gamma (\tau, \vec{\sigma}) &=&
\partial_r \pi^r (\tau, \vec{\sigma})
+ e \stackrel{o}{\bar{\psi}} (\tau, \vec{\sigma}) \gamma^{\bar{o}}
\stackrel{o}{\psi} (\tau, \vec{\sigma}) \approx 0, \nonumber \\
\tilde{H}^\mu (\tau) &=&p^{\mu}_s-[u^{\mu}(p_s) H_{rel}(\tau )+\epsilon^{\mu}
_r(u(p_s)) H_{p r}(\tau )]=\nonumber \\
&=&u^\mu (p_s) H (\tau) +
\epsilon^\mu_r ( u(p_s)) H_{pr} (\tau) \approx 0 ,\nonumber \\
&&{}\nonumber \\
&&or\nonumber \\
&&{}\nonumber \\
H (\tau) &=& \eta_s\sqrt{p^2_s}-H_{rel}(\tau )=\nonumber \\ 
&=&\eta_s \sqrt{p^2_s} -
\int{d^3\sigma \Big[ \frac i2 \Big( \stackrel{o}{\bar{\psi}} 
(\tau, \vec{\sigma})
\gamma_r \partial_r \stackrel{o}{\psi} (\tau, \vec{\sigma})} + \nonumber \\
&-& \partial_r \stackrel{o}{\bar{\psi}} (\tau, \vec{\sigma})
\gamma_r \stackrel{o}{\psi} (\tau, \vec{\sigma}) \Big) +
 m \stackrel{o}{\bar{\psi}} (\tau, \vec{\sigma})
 \stackrel{o}{\psi} (\tau, \vec{\sigma}) + \nonumber \\
&+& \frac12 \Big( \vec{\pi}^2 + \vec{B}^2 \Big) (\tau, \vec{\sigma}) +
 e A_r (\tau, \vec{\sigma}) \stackrel{o}{\bar{\psi}} (\tau, \vec{\sigma})
\gamma_r \stackrel{o}{\psi} (\tau, \vec{\sigma}) \Big]\approx 0,  \nonumber \\
H_{pr} (\tau) &=&
\int{d^3\sigma \Big[ \frac i2 \Big( \stackrel{o}{\bar{\psi}} (\tau, 
\vec{\sigma})
\gamma^{\bar{o}} \partial_r \stackrel{o}{\psi} (\tau, \vec{\sigma}) -
\partial_r \stackrel{o}{\bar{\psi}} (\tau, \vec{\sigma})
\gamma^{\bar{o}} \stackrel{o}{\psi} (\tau, \vec{\sigma}) \Big)} + \nonumber \\
&+& \Big( \vec{\pi} \wedge \vec{B} \Big)_r (\tau, \vec{\sigma}) +
e A_r (\tau, \vec{\sigma}) \stackrel{o}{\bar{\psi}} (\tau, \vec{\sigma})
\gamma^{\bar{o}} \stackrel{o}{\psi} (\tau, \vec{\sigma}) \Big] \approx 0, 
\nonumber \\
&&{}\nonumber \\
\{ \tilde{H}^\mu (\tau) , \tilde{H}^\nu (\tau) \}^{\ast \ast}_D &=&
\int{d^3\sigma \Big[  \Big( u^\mu (p_s)\epsilon^\nu_r ( u(p_s)) -
u^\nu (p_s)\epsilon^\mu_r ( u(p_s)) \Big) \pi^r (\tau, \vec{\sigma})} + 
\nonumber \\
&-& \epsilon^\mu_r ( u(p_s)) F_{rs}(\tau, \vec{\sigma})
\epsilon^\nu_s ( u(p_s)) \Big] \Gamma (\tau, \vec{\sigma}) \approx 0 ,
\nonumber \\
\{ \Gamma (\tau, \vec{\sigma}) , \tilde{H}^\mu (\tau) \}^{\ast \ast}_D &=&
\{ \Gamma (\tau, \vec{\sigma}) ,
\Gamma (\tau, \vec{\sigma}^\prime)\}^{\ast \ast}_D =
\{ \Gamma (\tau, \vec{\sigma}) ,
\pi^\tau (\tau, \vec{\sigma}^\prime)\}^{\ast \ast}_D = \nonumber \\
&=& \{ \pi^\tau (\tau, \vec{\sigma}) , \tilde{H}^\mu (\tau) \}^{\ast \ast}_D
= \{ \pi^\tau (\tau, \vec{\sigma}) ,
\pi^\tau (\tau, \vec{\sigma}^\prime) \} = 0 .
\label{IV14}
\end{eqnarray}

The constraints ${\vec H}_p(\tau )\approx 0$ identify the Wigner hyperplane
$\Sigma_{\tau W}$ with the intrinsic rest frame (vanishing of the total Wigner 
spin 1 ${}$ 3-momentum of the isolated system) and say that the 3-coordinate
$\vec \sigma ={\vec x}_{com}$ [$x^{\mu}_{com}=z^{\mu}(\tau ,{\vec x}_{com})$]
of the center of mass of the isolated system on
$\Sigma_{\tau W}$ is a gauge variable, whose natural gauge-fixing is
${\vec x}_{com}\approx 0$ [so that it coincides with the origin of $\Sigma
_{\tau W}$: ${}x^{\mu}_s(\tau )=z^{\mu}(\tau ,\vec \sigma =0)$]. See
Ref.\cite{mate} for the definition of $x^{\mu}_{com}$ for the configurations of
the Klein-Gordon field. The
remaining constraint $H(\tau )\approx 0$ identifies $\epsilon_s=\eta_s
\sqrt{p^2_s}$ with the invariant mass $H_{rel}$ of the isolated system.

On $\Sigma_{\tau W}$ the Dirac Hamiltonian becomes

\begin{equation}
H_D = \lambda (\tau) H (\tau)
- \vec{\lambda} (\tau) \vec{H}_p (\tau) +
\int{d^3\sigma \Big[ \mu_\tau (\tau, \vec{\sigma}) \pi^\tau (\tau, \vec{\sigma})
- A_\tau (\tau, \vec{\sigma}) \Gamma (\tau, \vec{\sigma}) \Big]} ,
\label{IV15}
\end{equation}

\noindent so that ${\tilde x}^{\mu}_s$ has a velocity parallel to $p^{\mu}_s$
[$\dot{\tilde{x}}^\mu_s (\tau) =
\{ \tilde{x}^\mu_s , H_D \}^{\ast \ast}_D = -\lambda (\tau) u^\mu (p_s)$],
namely it has no classical zitterbewegung.

Since ${\tilde H}^{\mu\nu}(\tau )\equiv 0$ implies

\begin{eqnarray}
S^{\mu \nu} &=& S^{\mu \nu}_\psi +
\Big( \epsilon^\mu_r (u(p_s)) u^\nu(p_s) + \nonumber \\
&-& \epsilon^\nu_r (u(p_s)) u^\mu(p_s) \Big)
\int{d^3\sigma \sigma^r \Big[ \frac i2
\Big(\stackrel{o}{\bar{\psi}} (\tau, \vec{\sigma})
\gamma_s \partial_s \stackrel{o}{\psi} (\tau, \vec{\sigma})} + \nonumber \\
&-& \partial_s \stackrel{o}{\bar{\psi}} (\tau, \vec{\sigma})
\gamma_s \stackrel{o}{\psi} (\tau, \vec{\sigma}) \Big) +
 m \stackrel{o}{\bar{\psi}} (\tau, \vec{\sigma})
 \stackrel{o}{\psi} (\tau, \vec{\sigma}) + \nonumber \\
&+& \frac12 \Big( \vec{\pi}^2 + \vec{B}^2 \Big) (\tau, \vec{\sigma}) +
 e A_s (\tau, \vec{\sigma}) \stackrel{o}{\bar{\psi}} (\tau, \vec{\sigma})
\gamma_s \stackrel{o}{\psi} (\tau, \vec{\sigma}) \Big] + \nonumber \\
&-& \Big( \epsilon^\mu_r (u(p_s)) \epsilon^\nu_s(p_s) -
 \epsilon^\nu_r (u(p_s)) \epsilon_s^\mu(p_s) \Big)
\int{d^3\sigma \sigma^r \Big[
\frac i2 \Big( \stackrel{o}{\bar{\psi}} (\tau, \vec{\sigma})
\gamma^{\bar{o}} \partial_s \stackrel{o}{\psi} (\tau, \vec{\sigma})} + 
\nonumber \\
 &-& \partial_s \stackrel{o}{\bar{\psi}} (\tau, \vec{\sigma})
\gamma^{\bar{o}} \stackrel{o}{\psi} (\tau, \vec{\sigma}) \Big) +
 \Big( \vec{\pi} \wedge \vec{B} \Big)_s (\tau, \vec{\sigma}) +
e A_s (\tau, \vec{\sigma}) \stackrel{o}{\bar{\psi}} (\tau, \vec{\sigma})
\gamma^{\bar{o}} \stackrel{o}{\psi} (\tau, \vec{\sigma}) \Big] ,
\label{IV16} 
\end{eqnarray}

\noindent with

\begin{equation}
S^{\mu \nu}_\psi = \frac 14 L^\mu_{.\alpha} (p_s , \stackrel{o}p_s)
L^\nu_{.\beta} (p_s , \stackrel{o}p_s) \int{d^3\sigma
\stackrel{o}{\bar{\psi}} (\tau, \vec{\sigma})
\big[ \gamma^{\bar{o}}, \sigma^{\alpha \beta} \big]_+
\stackrel{o}{\psi} (\tau, \vec{\sigma})} ,
\label{IV17}
\end{equation}

\noindent we get the following expression for the rest-frame spin tensor

\begin{eqnarray}
\bar{S}^{AB} &=& \bar{S}^{AB}_\psi +
\Big[ \delta^A_r \delta^B_{\bar{o}} - \delta^B_r \delta^A_{\bar{o}} \Big]
\int{d^3\sigma \sigma^r \Big[ \frac i2
\Big(\stackrel{o}{\bar{\psi}} (\tau, \vec{\sigma})
\gamma_s \partial_s \stackrel{o}{\psi} (\tau, \vec{\sigma})} + \nonumber \\
&-& \partial_s \stackrel{o}{\bar{\psi}} (\tau, \vec{\sigma})
\gamma_s \stackrel{o}{\psi} (\tau, \vec{\sigma}) \Big) +
 m \stackrel{o}{\bar{\psi}} (\tau, \vec{\sigma})
 \stackrel{o}{\psi} (\tau, \vec{\sigma}) + \nonumber \\
&+& \frac12 \Big( \vec{\pi}^2 + \vec{B}^2 \Big) (\tau, \vec{\sigma}) +
 e A_s (\tau, \vec{\sigma}) \stackrel{o}{\bar{\psi}} (\tau, \vec{\sigma})
\gamma_s \stackrel{o}{\psi} (\tau, \vec{\sigma}) \Big] + \nonumber \\
&-& \Big[ \delta^A_r \delta^B_s - \delta^B_r \delta^A_s \Big]
\int{d^3\sigma \sigma^r \Big[
\frac i2 \Big( \stackrel{o}{\bar{\psi}} (\tau, \vec{\sigma})
\gamma^{\bar{o}} \partial_s \stackrel{o}{\psi} (\tau, \vec{\sigma})} + 
\nonumber \\
 &-& \partial_s \stackrel{o}{\bar{\psi}} (\tau, \vec{\sigma})
\gamma^{\bar{o}} \stackrel{o}{\psi} (\tau, \vec{\sigma}) \Big) +
 \Big( \vec{\pi} \wedge \vec{B} \Big)_s (\tau, \vec{\sigma}) +
e A_s (\tau, \vec{\sigma}) \stackrel{o}{\bar{\psi}} (\tau, \vec{\sigma})
\gamma^{\bar{o}} \stackrel{o}{\psi} (\tau, \vec{\sigma}) \Big] ,
\label{IV18}
\end{eqnarray}

\noindent where

\begin{equation}
\bar{S}^{AB}_\psi = \frac 14  \int{d^3\sigma
\stackrel{o}{\bar{\psi}} (\tau, \vec{\sigma})
\big[ \gamma^{\bar{o}}, \sigma^{AB} \big]_+
\stackrel{o}{\psi} (\tau, \vec{\sigma})}
\label{IV19}
\end{equation}

\noindent is the component connected with the Dirac field.

On $\Sigma_{\tau W}$ the Poincar\'e generators are

\begin{eqnarray}
p_s^\mu,&&~~~~~J^{\mu \nu} =  \hat{x}^\mu_s p_s^\nu - \hat{x}^\nu_s p_s^\mu +
\tilde{S}^{\mu \nu},\nonumber \\
&&{}\nonumber \\
\tilde{S}^{\mu \nu} &\equiv & \tilde{S}_s^{\mu \nu}
 + \tilde{S}_\xi^{\mu \nu}, \nonumber \\
\tilde{S}^{0i} &=& -\frac{\delta^{ir} \bar{S}^{rs} p_s^s}
{p_s^0 + \eta_s \sqrt{p_s^2}},~~~~~\tilde{S}^{ij} = \delta^{ir} \delta^{js}
\bar{S}^{rs} ,
\label{IV20}
\end{eqnarray}

\noindent because one can express ${\tilde S}^{\mu\nu}$ in terms of ${\bar S}
^{AB}=\epsilon^A_{\mu}(u(p_s))\epsilon (u(p_s)) S^{\mu\nu}$. Only the Thomas
spin $\bar{S}_r = \frac12 \epsilon_{rst} \bar{S}_{st}$ contributes to them.
This is the universal realization of the Poincar\'e generators associated with 
the rest-frame Wigner-covariant instant form of the dynamics.

The electric charge takes the form

\begin{equation}
Q= - e \int{d^3\sigma
\stackrel{o}{\bar{\psi}} (\tau, \vec{\sigma}) \gamma^{\bar{o}}
\stackrel{o}{\psi} (\tau, \vec{\sigma})},
\label{IV21}
\end{equation}

\noindent and is the weak Noether charge of the conserved 4-current
$j^A (\tau, \vec{\sigma}) = - e \stackrel{o}{\bar{\psi}} (\tau, \vec{\sigma})
 \gamma^A \stackrel{o}{\psi} (\tau, \vec{\sigma})$,
$\partial_A j^A (\tau, \vec{\sigma})\, {\buildrel \circ \over =}\, 0$.

Therefore, the rest-frame Wigner-covariant instant form of the system Dirac
plus electromagnetic fields (pseudoclassical electrodynamics) formally coincides
with the standard noncovariant Hamiltonian formulation of the system on the 
hyperplanes $z^o(\tau ,\vec \sigma )=const.$ : only the covariance properties 
of the objects are different and, moreover, there are the first class 
constraints ${\vec H}_p(\tau )\approx 0$ defining the rest frame.

\vfill\eject

\section{Dirac's Observables and Equations of Motion.}

As shown in Ref.\cite{lusa}, the Dirac observables of the electromagnetic field 
are the transverse quantities ${\vec A} (\tau, \vec{\sigma})$,
${\vec \pi}(\tau, \vec{\sigma})$, defined by the decomposition

\begin{eqnarray}
\vec A (\tau, \vec{\sigma}) &=& \vec \partial \eta (\tau, \vec{\sigma}) +
{\vec A}_{\perp } (\tau, \vec{\sigma}), \nonumber \\
\vec \pi (\tau, \vec{\sigma}) &=& {\vec \pi}_{\perp} (\tau, \vec{\sigma}) +
\frac{\vec \partial}{\triangle_\sigma}
\Big[\Gamma (\tau, \vec{\sigma}) - \sum_{i=1}^N Q_i(\tau)
\delta^3(\vec{\sigma} - \vec{\eta}_i(\tau)) \Big] , \nonumber \\
\eta (\tau, \vec{\sigma}) &=& - \frac{\vec{\partial}}{\triangle_\sigma}
\cdot \vec{A} (\tau, \vec{\sigma}),
\label{V1}
\end{eqnarray}

\noindent while the gauge variables are $A_\tau (\tau, \vec{\sigma})$ and
$\eta (\tau, \vec{\sigma})$, which are conjugate to the first class constraints
$\pi^{\tau}(\tau, \vec{\sigma})\approx 0$, $\Gamma (\tau, \vec{\sigma})\approx
0$.

Since we have

\begin{equation}
\{ {\buildrel \circ \over \psi}(\tau ,\vec \sigma ),\Gamma (\tau 
,{\vec \sigma}^{'}) \}^{**}_D =-ie \gamma^{\tau}\, \psi (\tau ,\vec \sigma )
\delta^3(\vec \sigma -{\vec \sigma}^{'}),
\label{V2}
\end{equation}

\noindent we see that the Dirac field is not gauge invariant. Like in Ref.
\cite{lusa}, it turns out that its Dirac observables are

\begin{eqnarray}
\check{\stackrel{o}\psi} (\tau, \vec{\sigma}) &=&
e^{- i e \eta (\tau, \vec{\sigma}) } \stackrel{o}\psi (\tau, \vec{\sigma}) ,
\nonumber \\
\check{\stackrel{o}{\bar{\psi}}} (\tau, \vec{\sigma}) &=&
\stackrel{o}{\bar{\psi}} (\tau, \vec{\sigma})
e^{ i e \eta (\tau, \vec{\sigma}) } ,\nonumber \\
&&\{ \check{\stackrel{o}\psi^a} (\tau, \vec{\sigma}) ,
\check{\stackrel{o}{\bar{\psi}}}_b (\tau, \vec{\sigma}^\prime) \}^{\ast \ast}_D
= - i (\gamma^{\bar{o}})^a_{.b}
\delta^3 (\vec{\sigma} - \vec{\sigma}^\prime) ,
\label{V3}
\end{eqnarray}

\noindent representing Dirac fields dressed with a Coulomb cloud.

By using Eqs.(\ref{V1}) we get

\begin{eqnarray}
\int{d^3\sigma \vec{\pi}^2 (\tau, \vec{\sigma})} &=&
\int{d^3\sigma \vec{\pi}^2_\perp (\tau, \vec{\sigma})} + \nonumber \\
&+& e^2 \int{\int{ d^3\sigma d^3\sigma^\prime
\frac { \Big[ \stackrel{o}{\bar{\psi}} (\tau, \vec{\sigma}) \gamma^{\bar{o}}
\stackrel{o}\psi (\tau, \vec{\sigma}) \Big]
\Big[ \stackrel{o}{\bar{\psi}} (\tau, \vec{\sigma}^\prime) \gamma^{\bar{o}}
\stackrel{o}\psi (\tau, \vec{\sigma}^\prime) \Big]}
{4\pi \mid \vec{\sigma} - \vec{\sigma}^\prime \mid}}} = \nonumber \\
&=& \int{d^3\sigma \vec{\pi}^2_\perp (\tau, \vec{\sigma})} + \nonumber \\
 &+& e^2 \int{\int{ d^3\sigma d^3\sigma^\prime
\frac { \Big[ \check{\stackrel{o}{\bar{\psi}}} (\tau, \vec{\sigma}) 
\gamma^{\bar{o}}
\check{\stackrel{o}\psi} (\tau, \vec{\sigma}) \Big]
\Big[ \check{\stackrel{o}{\bar{\psi}}} (\tau, \vec{\sigma}^\prime)
\gamma^{\bar{o}}
\check{\stackrel{o}\psi} (\tau, \vec{\sigma}^\prime) \Big]}
{4\pi \mid \vec{\sigma} - \vec{\sigma}^\prime \mid}}},
\label{V4} 
\end{eqnarray}

\noindent so that in the Coulomb gauge, $A_{\tau}(\tau, \vec{\sigma})=\pi^{\tau}
(\tau, \vec{\sigma})=\eta (\tau, \vec{\sigma})=\Gamma (\tau, \vec{\sigma})=0$,
we have [$\epsilon_s=\eta_s\sqrt{p^2_s}$]

\begin{eqnarray}
H(\tau) &=& \epsilon_s-H_{rel}(\tau ) =\nonumber \\
&=&\eta_s \sqrt{p_s^2} - \int{d^3\sigma\Big[ \frac i2
\Big(\check{\stackrel{o}{\bar{\psi}}} (\tau, \vec{\sigma})
\gamma_r \partial_r \check{\stackrel{o}{\psi}} (\tau, \vec{\sigma})} + 
\nonumber \\
&-& \partial_r \check{\stackrel{o}{\bar{\psi}}} (\tau, \vec{\sigma})
\gamma_r \check{\stackrel{o}{\psi}} (\tau, \vec{\sigma}) \Big) +
 m \check{\stackrel{o}{\bar{\psi}}} (\tau, \vec{\sigma})
 \check{\stackrel{o}{\psi}} (\tau, \vec{\sigma}) + \nonumber \\
&+& \frac12 \Big( \vec{\pi}_\perp^2 + \vec{B}^2 \Big) (\tau, \vec{\sigma}) +
 e A_{\perp r} (\tau, \vec{\sigma})
 \check{\stackrel{o}{\bar{\psi}}} (\tau, \vec{\sigma})
\gamma_r \check{\stackrel{o}{\psi}} (\tau, \vec{\sigma}) + \nonumber \\
&+& \frac{e^2}{2}  \check{\stackrel{o}{\bar{\psi}}} (\tau, \vec{\sigma})
\gamma^{\bar{o}} \check{\stackrel{o}{\psi}} (\tau, \vec{\sigma})
\int{d^3\sigma^\prime
\frac{\check{\stackrel{o}{\bar{\psi}}} (\tau, \vec{\sigma}^\prime)
\gamma^{\bar{o}}
\check{\stackrel{o}\psi} (\tau, \vec{\sigma}^\prime)}
{4\pi \mid \vec{\sigma} - \vec{\sigma}^\prime \mid}} \Big] \approx 0.
\label{V5}
\end{eqnarray}

Like in Ref.\cite{lusa}, the last term is the nonrenormalizable Coulomb self 
interaction of the Dirac field in the rest frame. The constraints identifying 
the rest frame take the form expected in an instant form of dynamics [they do 
not depend on the interaction]

\begin{equation}
H_{pr}(\tau) = \int{d^3\sigma \Big[ \frac 12
\Big(\check{\stackrel{o}{\bar{\psi}}} (\tau, \vec{\sigma})
\gamma^{\bar{o}} \partial_r \check{\stackrel{o}{\psi}} (\tau, \vec{\sigma})}
- \partial_r \check{\stackrel{o}{\bar{\psi}}} (\tau, \vec{\sigma})
\gamma^{\bar{o}} \check{\stackrel{o}{\psi}} (\tau, \vec{\sigma}) \Big) +
\Big( \vec{\pi}_\perp \wedge \vec{B} \Big)_r (\tau, \vec{\sigma})
\approx 0.
\label{V6}
\end{equation}

By doing the canonical transformation\cite{longhi} from the variables
$\tilde{x}^\mu_s (\tau)$ e $p^\mu_s$, to the new ones

\begin{eqnarray}
T_s &=& \frac{{p_s}_\mu \tilde{x}^\mu_s}{\eta_s \sqrt{p^2_s}}
=  \frac{{p_s}_\mu x^\mu_s}{\eta_s \sqrt{p^2_s}}, ~~~~~
\epsilon_s = \eta_s \sqrt{p^2_s}, \nonumber \\
\vec{z}_s &=& \eta_s \sqrt{p^2_s} \Big( \vec{\tilde{x}}_s -
\frac{\vec{p}_s}{p^0} \tilde{x}^0 \Big), ~~~~~
\vec{k}_s = \frac{\vec{p}_s} {\eta_s \sqrt{p^2_s}} ,\nonumber \\
&&\{ T_s , \epsilon_s \}^{\ast \ast}_D = - 1,\quad\quad
\{ z^i_s , k^j_s \}^{\ast \ast}_D = \delta^{ij},
\label{V7}
\end{eqnarray}

\noindent we arrive at a canonical basis with the rest-frame Lorentz-scalar 
time $T_s$ and with the canonical noncovariant 3-variable ${\vec z}_s$ 
[replacing the origin of $\Sigma_{\tau W}$] with the same covariance of the Newton-Wigner
position operator.

By adding the gauge-fixing $T_s - \tau \approx 0$ [which identifies the 
rest-frame time $T_s$ with the parameter $\tau$ of the foliation of Minkowski 
spacetime with the Wigner hyperplanes associated with the isolated system], 
whose time constancy implies $\lambda(\tau) = -1$, we get the Dirac Hamiltonian

\begin{equation}
\hat{H}_D = H_{rel}(\tau) - \vec{\lambda}(\tau) \cdot \vec{H}_p(\tau),
\label{V8}
\end{equation}

\noindent where

\begin{eqnarray}
H_{rel} &=& \int{d^3\sigma \Big[ \frac i2 \Big(
\check{\stackrel{o}{\bar{\psi}}} (\tau, \vec{\sigma})
\gamma_r \partial_r \check{\stackrel{o}{\psi}} (\tau, \vec{\sigma})}
- \partial_r \check{\stackrel{o}{\bar{\psi}}} (\tau, \vec{\sigma})
\gamma_r \check{\stackrel{o}{\psi}} (\tau, \vec{\sigma}) \Big)+ \nonumber \\
&+& m \check{\stackrel{o}{\bar{\psi}}} (\tau, \vec{\sigma})
\check{\stackrel{o}{\psi}} (\tau, \vec{\sigma})
+ \frac 12 (\vec{\pi}^2_\perp + \vec{B}^2) (\tau, \vec{\sigma})\Big]+
\nonumber \\
&+& \frac{e^2}{2} \int{\int{d^3\sigma d^3\sigma^\prime
\frac{\big[ \check{\stackrel{o}{\bar{\psi}}} (\tau, \vec{\sigma})
\gamma^{\bar{o}} \check{\stackrel{o}{\psi}} (\tau, \vec{\sigma}) \big]
\big[ \check{\stackrel{o}{\bar{\psi}}} (\tau, \vec{\sigma}^\prime)
\gamma^{\bar{o}} \check{\stackrel{o}{\psi}} (\tau, \vec{\sigma}^\prime) \big]}
{ 4\pi \mid \vec{\sigma} - \vec{\sigma}^\prime \mid} }} + \nonumber \\
&+& e \int{d^3\sigma A_{r \perp}(\tau, \vec{\sigma})
\check{\stackrel{o}{\bar{\psi}}} (\tau, \vec{\sigma})
\gamma_r \check{\stackrel{o}{\psi}} (\tau, \vec{\sigma})}.
\label{V9}
\end{eqnarray}

In the gauge $\vec \lambda (\tau )=0$, the Dirac field has the following
Hamilton equation

\begin{eqnarray}
\partial_\tau \check{\stackrel{o}\psi} (\tau, \vec{\sigma})
\, &{\buildrel \circ \over =}\,&
\{ \check{\stackrel{o}\psi} (\tau, \vec{\sigma}) ,
\hat{H}_D \}^{\ast \ast}_D =  \nonumber \\
&=& \gamma^{\bar{o}} \gamma_r \Big[ \partial_r -
i e A_{\perp r} (\tau, \vec{\sigma}) \Big]
\check{\stackrel{o}\psi} (\tau, \vec{\sigma}) - i m  \gamma^{\bar{o}}
\check{\stackrel{o}\psi} (\tau, \vec{\sigma}) + \nonumber \\
&-& i e^2 \int{d^3\sigma^\prime
\frac{\check{\stackrel{o}{\bar{\psi}}} (\tau, \vec{\sigma}^\prime)
 \gamma^{\bar{o}}
\check{\stackrel{o}\psi} (\tau, \vec{\sigma}^\prime) }
{4 \pi \mid \vec{\sigma} - \vec{\sigma}^\prime \mid }
\check{\stackrel{o}\psi} (\tau, \vec{\sigma}) },
\label{V10}
\end{eqnarray}

\noindent which can be rewritten in the standard form

\begin{equation}
\{ i \gamma^A [\partial_A - i e \tilde{A}_A (\tau , \vec{\sigma}) ] - m \}
\check{\stackrel{o}\psi} (\tau, \vec{\sigma})\, {\buildrel \circ \over =}\, 0,
\label{V11}
\end{equation}

\noindent with

\begin{eqnarray}
\tilde{A}_r (\tau , \vec{\sigma}) &\equiv & A_{\perp r} (\tau , \vec{\sigma}),
\nonumber \\
\tilde{A}_{\bar{o}} (\tau , \vec{\sigma}) &\equiv &
- e \int{d^3\sigma^\prime
\frac{\check{\stackrel{o}{\bar{\psi}}} (\tau, \vec{\sigma}^\prime)
 \gamma^{\bar{o}}
\check{\stackrel{o}\psi} (\tau, \vec{\sigma}^\prime) }
{4 \pi \mid \vec{\sigma} - \vec{\sigma}^\prime \mid } } =
\tilde{A}_{\bar{o}}^\dag (\tau , \vec{\sigma}) .
\label{V12}
\end{eqnarray}

\noindent Analogously we get

\begin{equation}
\check{\stackrel{o}{\bar{\psi}}} (\tau, \vec{\sigma})
 \{ -i [\stackrel{\leftarrow}{\partial_A} +
i e \tilde{A}_A (\tau , \vec{\sigma}) ] \gamma^A - m \}\, {\buildrel \circ 
\over =}\, 0  .
\label{V13}
\end{equation}

Eqs.(\ref{V11}) and (\ref{V12}) are nonlocal and nonlinear due to the 
reduction to the rest-frame Coulomb gauge.

For the transverse electromagnetic fields $\vec{A}_\perp (\tau , \vec{\sigma})$,
$\vec{\pi}_\perp (\tau , \vec{\sigma})$ we get the Hamilton equations

\begin{eqnarray}
\partial_\tau A^r_\perp (\tau , \vec{\sigma})\, &{\buildrel \circ \over =}\,&
\{ A^r_\perp (\tau , \vec{\sigma}), \hat{H}_D \}^{\ast \ast}_D
= - \pi^r_\perp (\tau , \vec{\sigma}),\nonumber  \\
\partial_\tau \pi^r_\perp (\tau , \vec{\sigma})\, &{\buildrel \circ \over =}\,&
\{ \pi^r_\perp (\tau , \vec{\sigma}), \hat{H}_D \}^{\ast \ast}_D
= \triangle_\sigma A^r_\perp (\tau , \vec{\sigma}) + \nonumber \\
&+& e P^{rs} (\vec{\sigma}) \Big[
\check{\stackrel{o}{\bar{\psi}}} (\tau, \vec{\sigma})
 \gamma^s
\check{\stackrel{o}\psi} (\tau, \vec{\sigma})  \Big] ,
\label{V14}
\end{eqnarray}

\noindent implying the equation

\begin{equation}
{\Box}_\sigma A^r_\perp (\tau , \vec{\sigma})\, {\buildrel \circ \over =}\,
P^{rs} (\vec{\sigma}) j^s (\tau , \vec{\sigma}) \equiv
j^r_\perp (\tau , \vec{\sigma}),
\label{V15}
\end{equation}

\noindent with

\begin{eqnarray}
{\Box}_\sigma &\equiv & \partial_A \partial^A, \nonumber \\
P^{rs} (\vec{\sigma}) &\equiv & \delta^{rs} +
\frac{\partial^r \partial^s}{\triangle_\sigma}, \nonumber \\
j^s (\tau , \vec{\sigma}) &=& - e
\check{\stackrel{o}{\bar{\psi}}} (\tau, \vec{\sigma}) \gamma^s
\check{\stackrel{o}\psi} (\tau, \vec{\sigma}).
\label{V16}
\end{eqnarray}

The wave equation is actually an integrodifferential equation due to the
projector apearing in the transverse fermionic Wigner spin 1 ${}$ 3-current.

\vfill\eject

\section{Conclusions}

In this paper we have given the formulation of Grassmann-valued Dirac fields
plus the electromagnetic field on spacelike hypersurfaces in Minkowski
spacetime. Then, we have done the canonical reduction of this
pseudoclassical basis of QED to the rest-frame Wigner-covariant Coulomb gauge,
finding a canonical basis of Dirac's observables in the rest-frame
Wigner-covariant instant form of dynamics. What has still to be done is the
extension of the theory to include tetrad gravity.

These results are valid for massive Dirac fields and for all the massive
[$P^2 > 0$] configurations of massless Dirac fields. They also hold for the
massive configurations of chiral fields simply by replacing everywhere $\psi$ 
with $\psi_{\pm}={1\over 2}(1\pm \gamma_5)\psi$. 
Instead, the massless [$P^2=0$] 
and infrared [$P^{\mu}=0$] configurations of either massless or chiral fermion
fields have to be treated separately, because they require the reformulation
of the front form of dynamics in the instant form like for massless spinning
particles (this problem will be studied in a future paper).

Therefore, these results open the path to the reformulation and the canonical
reduction to a rest-frame Wigner-covariant Coulomb gauge of the
SU(3)xSU(2)xU(1) standard model, which will be studied elsewhere.

However, the rest-frame Wigner-covariant instant form evidentiates that
there is a nontrivial problem in the description of fermion fields: 
Eq.(\ref{IV14}) shows that the 
geometrical Minkowski 4-vector $p^{\mu}_s$ normal to the Wigner hyperplane
$\Sigma_{\tau W}$ is the sum of an even ``bosonic" part determined by the 
electromagnetic field plus a ``fermionic" part bilinear in the Grassmann-valued 
Dirac fields. In absence of the electromagnetic field, $p^{\mu}_s$ would weakly
become a bilinear in Grassmann variables, but this is inconsistent from both a 
geometrical point of view and from an algebraic one [one can neither define
$\sqrt{p^2_s}$, nor divide by it to get the standard Wigner boost due to the 
nilpotent character of Grassmann variables]. It seems that the Wigner 
hyperplanes and the rest-frame description of an isolated pseudoclassical Dirac 
field is impossible due to the necessity of having it Grassmann-valued as it 
is required to evaluate the fermionic path integral.

But this does not sound reasonable. Something is missing. In Ref.\cite{big},
we have seen that a spinning particle with a definite sign of the energy 
(but this holds also for the ordinary spinning particles) is described by a 
fibration: there is a scalar particle (the tracer of the Minkowski worldline, 
i.e. the path of the electric current if the spinning particle is charged) with 
a Grassmann fiber over it describing the spin structure and this is 
particularly evident in its supersymmetric description with the superfield 
$X^{\mu}=x^{\mu}+\theta \xi^{\mu}$ \cite{sp1} [$\xi^{\mu}$ are the Grassmann
variables for the spin description, which go into the Dirac matrices after
quantization]. As said in Ref.\cite{big}, one expects that this fibered 
structure should survive in first quantization: the electron wave function is
expected to be some kind of scalar superfield $\phi_D+\theta \psi_D$, where 
$\phi_D$ is a charged Klein-Gordon wave function (restricted to its lower 
level, with the same degrees of freedom of a scalar particle, to avoid the
introduction of 
spurious levels in bound state spectra) and $\psi_D$ is the ordinary Dirac 
wave function [$\psi_D$ should live in a Clifford fiber over $\phi_D$]. This
superfield does not describe a supersymmetric multiplet $\{ selectron,\,
electron \}$, but the fibration associated with the spin structure [see Ref.
\cite{g5} for the description of the pseudoclassical photon as a ray of light 
with a Grassmann fibration describing the spin structure, namely the light 
polarization]. The obstacle to arrive at this point of view is that the
first class constraint $p^2-m^2\approx 0$ has always been quantized in a form
which corresponds to the square of the Dirac equation so that it acts on Dirac
wave functions. But at the classical level it describes the bosonic scalar
particle tracing the worldline so that it has to be quantized so to act on
Klein-Gordon wave functions. The double role of this constraint is to give
a consistency between what happens on the base (Minkowski spacetime) and what
happens on the fiber.

Therefore, one expects that also Dirac fields should be replaced by a fibration 
$\varphi +\theta \psi$ describing the spin structure with a pair $(\varphi 
,\psi )$, where $\psi$ is the Grassmann-valued Dirac field and $\varphi$ a 
charged Klein-Gordon field suitably restricted.

We shall study this problem elsewhere. A guide to its solution will be to find
the way  to extract the rest-frame constraints of the spinning particle of 
Ref.\cite{big} from the fermion field theory, in the same way in which the 
analogous constraints for a 
scalar particle were extracted by the rest-frame description of scalar 
electrodynamics in Ref.\cite{lus2} by using the Feshbach-Villars 
transformation\cite{fv}; now both Feshbach-Villars and Foldy-Wouthuysen 
\cite{fw} transformations will be needed. The other ingredient will be the 
individuation of the subspace of Klein-Gordon configurations with the same
degrees of freedom of a scalar particle [monopole configurations], by
using the technology developed to study the canonical separation of the
center-of-mass degrees of freedom from the relative ones for Klein-Gordon
fields\cite{mate}.

Let us remark that the same problems are present in the nonrelativistic
description of fermion fileds [Grassmann-valued Dirac spinors are replaced
by Grassmann-valued Pauli spinors], needed for the theory of Cooper pairs
in superconductivity.

The problem of quantization of isolated systems in these rest-frame
Wigner-covariant Coulomb gauges is an open problem [see Refs.\cite{cou,lav}
for what is known on the quantization in the Coulomb gauge], which will be
the subject of future investigations trying to use the M$\o$ller radius
$\rho =\sqrt{-W^2}/P^2=|{\vec {\bar S}}^2|/\sqrt{P^2}$ as a physical
ultraviolet cutoff \cite{re}.

\vfill\eject

\appendix

\section{Foliations of Minkowski Spacetime with Families of Spacelike
Hypersurfaces}

Let us  review some preliminary results from Refs.\cite{lus1,lus2}
needed in the description of physical systems on spacelike hypersurfaces.

Let $\lbrace \Sigma_{\tau}\rbrace$ be a one-parameter family of spacelike
hypersurfaces foliating Minkowski spacetime $M^4$ and giving a 3+1 decomposition
of it. At fixed $\tau$, let 
$z^{\mu}(\tau ,\vec \sigma )$ be the coordinates of the points on $\Sigma
_{\tau }$ in $M^4$, $\lbrace \vec \sigma \rbrace$ a system of coordinates on
$\Sigma_{\tau}$. If $\sigma^{\check A}=(\sigma^{\tau}=\tau ;\vec \sigma 
=\lbrace \sigma^{\check r}\rbrace)$ [the notation ${\check A}=(\tau ,
{\check r})$ with ${\check r}=1,2,3$ will be used; note that ${\check A}=
\tau$ and ${\check A}={\check r}=1,2,3$ are Lorentz-scalar indices] and 
$\partial_{\check A}=\partial /\partial \sigma^{\check A}$, 
one can define the vierbeins

\begin{equation}
z^{\mu}_{\check A}(\tau ,\vec \sigma )=\partial_{\check A}z^{\mu}(\tau ,\vec 
\sigma ),\quad\quad
\partial_{\check B}z^{\mu}_{\check A}-\partial_{\check A}z^{\mu}_{\check B}=0,
\label {a1}
\end{equation}

\noindent so that the metric on $\Sigma_{\tau}$ is

\begin{eqnarray}
&&g_{{\check A}{\check B}}(\tau ,\vec \sigma )=z^{\mu}_{\check A}(\tau ,\vec 
\sigma )\eta_{\mu\nu}z^{\nu}_{\check B}(\tau ,\vec \sigma ),\quad\quad 
g_{\tau\tau}(\tau ,\vec \sigma ) > 0,\nonumber \\
&&g(\tau ,\vec \sigma )=-det\, ||\, g_{{\check A}{\check B}}(\tau ,\vec 
\sigma )\, || ={(det\, ||\, z^{\mu}_{\check A}(\tau ,\vec \sigma )\, ||)}^2,
\nonumber \\
&&\gamma (\tau ,\vec \sigma )=-det\, ||\, g_{{\check r}{\check s}}(\tau ,\vec 
\sigma )\, ||.
\label{a2}
\end{eqnarray}

If $\gamma^{{\check r}{\check s}}(\tau ,\vec \sigma )$ is the inverse of the 
3-metric $g_{{\check r}{\check s}}(\tau ,\vec \sigma )$ [$\gamma^{{\check r}
{\check u}}(\tau ,\vec \sigma )g_{{\check u}{\check s}}(\tau ,\vec 
\sigma )=\delta^{\check r}_{\check s}$], the inverse $g^{{\check A}{\check B}}
(\tau ,\vec \sigma )$ of $g_{{\check A}{\check B}}(\tau ,\vec \sigma )$ 
[$g^{{\check A}{\check C}}(\tau ,\vec \sigma )g_{{\check c}{\check b}}(\tau ,
\vec \sigma )=\delta^{\check A}_{\check B}$] is given by

\begin{eqnarray}
&&g^{\tau\tau}(\tau ,\vec \sigma )={{\gamma (\tau ,\vec \sigma )}\over
{g(\tau ,\vec \sigma )}},\nonumber \\
&&g^{\tau {\check r}}(\tau ,\vec \sigma )=-[{{\gamma}\over g} g_{\tau {\check 
u}}\gamma^{{\check u}{\check r}}](\tau ,\vec \sigma ),\nonumber \\
&&g^{{\check r}{\check s}}(\tau ,\vec \sigma )=\gamma^{{\check r}{\check s}}
(\tau ,\vec \sigma )+[{{\gamma}\over g}g_{\tau {\check u}}g_{\tau {\check v}}
\gamma^{{\check u}{\check r}}\gamma^{{\check v}{\check s}}](\tau ,\vec \sigma ),
\label{a3}
\end{eqnarray}

\noindent so that $1=g^{\tau {\check C}}(\tau ,\vec \sigma )g_{{\check C}\tau}
(\tau ,\vec \sigma )$ is equivalent to

\begin{equation}
{{g(\tau ,\vec \sigma )}\over {\gamma (\tau ,\vec \sigma )}}=g_{\tau\tau}
(\tau ,\vec \sigma )-\gamma^{{\check r}{\check s}}(\tau ,\vec \sigma )
g_{\tau {\check r}}(\tau ,\vec \sigma )g_{\tau {\check s}}(\tau ,\vec \sigma ).
\label{a4}
\end{equation}

We have

\begin{equation}
z^{\mu}_{\tau}(\tau ,\vec \sigma )=(\sqrt{ {g\over {\gamma}} }l^{\mu}+
g_{\tau {\check r}}\gamma^{{\check r}{\check s}}z^{\mu}_{\check s})(\tau ,
\vec \sigma ),
\label{a5}
\end{equation}

\noindent and

\begin{eqnarray}
\eta^{\mu\nu}&=&z^{\mu}_{\check A}(\tau ,\vec \sigma )g^{{\check A}{\check B}}
(\tau ,\vec \sigma )z^{\nu}_{\check B}(\tau ,\vec \sigma )=\nonumber \\
&=&(l^{\mu}l^{\nu}+z^{\mu}_{\check r}\gamma^{{\check r}{\check s}}
z^{\nu}_{\check s})(\tau ,\vec \sigma ),
\label{a6}
\end{eqnarray}

\noindent where

\begin{eqnarray}
l^{\mu}(\tau ,\vec \sigma )&=&({1\over {\sqrt{\gamma}} }\epsilon^{\mu}{}_{\alpha
\beta\gamma}z^{\alpha}_{\check 1}z^{\beta}_{\check 2}z^{\gamma}_{\check 3})
(\tau ,\vec \sigma ),\nonumber \\
&&l^2(\tau ,\vec \sigma )=1,\quad\quad l_{\mu}(\tau ,\vec \sigma )z^{\mu}
_{\check r}(\tau ,\vec \sigma )=0,
\label{a7}
\end{eqnarray}

\noindent is the unit (future pointing) normal to $\Sigma_{\tau}$ at
$z^{\mu}(\tau ,\vec \sigma )$.

For the volume element in Minkowski spacetime we have

\begin{eqnarray}
d^4z&=&z^{\mu}_{\tau}(\tau ,\vec \sigma )d\tau d^3\Sigma_{\mu}=d\tau [z^{\mu}
_{\tau}(\tau ,\vec \sigma )l_{\mu}(\tau ,\vec \sigma )]\sqrt{\gamma
(\tau ,\vec \sigma )}d^3\sigma=\nonumber \\
&=&\sqrt{g(\tau ,\vec \sigma )} d\tau d^3\sigma.
\label{a8}
\end{eqnarray}

Let us remark that according to the geometrical approach of 
Ref.\cite{kuchar},one 
can use Eq.(\ref{a5}) in the form $z^{\mu}_{\tau}(\tau ,\vec \sigma )=N(\tau ,
\vec \sigma )l^{\mu}(\tau ,\vec \sigma )+N^{\check r}(\tau ,\vec \sigma )
z^{\mu}_{\check r}(\tau ,\vec \sigma )$, where $N=\sqrt{g/\gamma}=\sqrt{g
_{\tau\tau}-\gamma^{{\check r}{\check s}}g_{\tau{\check r}}g_{\tau{\check s}}}$ 
and $N^{\check r}=g_{\tau \check s}\gamma^{\check s\check r}$ are the 
standard lapse and shift functions, so that $g_{\tau \tau}=N^2+
g_{\check r\check s}N^{\check r}N^{\check s}, g_{\tau \check r}=
g_{\check r\check s}N^{\check s},
g^{\tau \tau}=N^{-2}, g^{\tau \check r}=-N^{\check r}/N^2, g^{\check r\check
s}=\gamma^{\check r\check s}+{{N^{\check r}N^{\check s}}\over {N^2}}$,
${{\partial}\over {\partial z^{\mu}_{\tau}}}=l_{\mu}\, {{\partial}\over
{\partial N}}+z_{{\check s}\mu}\gamma^{{\check s}{\check r}} {{\partial}\over
{\partial N^{\check r}}}$, $d^4z=N\sqrt{\gamma}d\tau d^3\sigma$.

The cotetrads $z^{\check A}_{\mu}(\tau ,\vec \sigma )$ have the expression

\begin{eqnarray}
&&z^{\tau}_{\mu}(\tau ,\vec \sigma )={1\over {N(\tau ,\vec \sigma )}}l_{\mu}
(\tau ,\vec \sigma ),\nonumber \\
&&z^{\check r}_{\mu}(\tau ,\vec \sigma )=-{{N^{\check r}(\tau ,\vec \sigma )}
\over {N(\tau ,\vec \sigma )}}l_{\mu}(\tau ,\vec \sigma )+\gamma^{\check 
r\check s}(\tau ,\vec \sigma )z_{\check s\mu}(\tau ,\vec \sigma ).
\label{a9}
\end{eqnarray}

The rest frame form of a timelike fourvector $p^{\mu}$ is $\stackrel
{\circ}{p}{}^{\mu}=\eta \sqrt{p^2} (1;\vec 0)= \eta^{\mu o}\eta \sqrt{p^2}$,
$\stackrel{\circ}{p}{}^2=p^2$, where $\eta =sign\, p^o$.
The standard Wigner boost transforming $\stackrel{\circ}{p}{}^{\mu}$ into
$p^{\mu}$ is

\begin{eqnarray}
L^{\mu}{}_{\nu}(p,\stackrel{\circ}{p})&=&\epsilon^{\mu}_{\nu}(u(p))=
\nonumber \\
&=&\eta^{\mu}_{\nu}+2{ {p^{\mu}{\stackrel{\circ}{p}}_{\nu}}\over {p^2}}-
{ {(p^{\mu}+{\stackrel{\circ}{p}}^{\mu})(p_{\nu}+{\stackrel{\circ}{p}}_{\nu})}
\over {p\cdot \stackrel{\circ}{p} +p^2} }=\nonumber \\
&=&\eta^{\mu}_{\nu}+2u^{\mu}(p)u_{\nu}(\stackrel{\circ}{p})-{ {(u^{\mu}(p)+
u^{\mu}(\stackrel{\circ}{p}))(u_{\nu}(p)+u_{\nu}(\stackrel{\circ}{p}))}
\over {1+u^o(p)} },\nonumber \\
&&{} \nonumber \\
\nu =0 &&\epsilon^{\mu}_o(u(p))=u^{\mu}(p)=p^{\mu}/\eta \sqrt{p^2}, \nonumber \\
\nu =r &&\epsilon^{\mu}_r(u(p))=(-u_r(p); \delta^i_r-{ {u^i(p)u_r(p)}\over
{1+u^o(p)} }).
\label{a10}
\end{eqnarray}

The inverse of $L^{\mu}{}_{\nu}(p,\stackrel{\circ}{p})$ is $L^{\mu}{}_{\nu}
(\stackrel{\circ}{p},p)$, the standard boost to the rest frame, defined by

\begin{equation}
L^{\mu}{}_{\nu}(\stackrel{\circ}{p},p)=L_{\nu}{}^{\mu}(p,\stackrel{\circ}{p})=
L^{\mu}{}_{\nu}(p,\stackrel{\circ}{p}){|}_{\vec p\rightarrow -\vec p}.
\label{a11}
\end{equation}

Therefore, we can define the following vierbeins [the $\epsilon^{\mu}_r(u(p))$'s
are also called polarization vectors; the indices r, s will be used for A=1,2,3
and $\bar o$ for A=0]

\begin{eqnarray}
&&\epsilon^{\mu}_A(u(p))=L^{\mu}{}_A(p,\stackrel{\circ}{p}),\nonumber \\
&&\epsilon^A_{\mu}(u(p))=L^A{}_{\mu}(\stackrel{\circ}{p},p)=\eta^{AB}\eta
_{\mu\nu}\epsilon^{\nu}_B(u(p)),\nonumber \\
&&{} \nonumber \\
&&\epsilon^{\bar o}_{\mu}(u(p))=\eta_{\mu\nu}\epsilon^{\nu}_o(u(p))=u_{\mu}(p),
\nonumber \\
&&\epsilon^r_{\mu}(u(p))=-\delta^{rs}\eta_{\mu\nu}\epsilon^{\nu}_r(u(p))=
(\delta^{rs}u_s(p);\delta^r_j-\delta^{rs}\delta_{jh}{{u^h(p)u_s(p)}\over
{1+u^o(p)} }),\nonumber \\
&&\epsilon^A_o(u(p))=u_A(p),
\label{a12}
\end{eqnarray}

\noindent which satisfy

\begin{eqnarray}
&&\epsilon^A_{\mu}(u(p))\epsilon^{\nu}_A(u(p))=\eta^{\mu}_{\nu},\nonumber \\
&&\epsilon^A_{\mu}(u(p))\epsilon^{\mu}_B(u(p))=\eta^A_B,\nonumber \\
&&\eta^{\mu\nu}=\epsilon^{\mu}_A(u(p))\eta^{AB}\epsilon^{\nu}_B(u(p))=u^{\mu}
(p)u^{\nu}(p)-\sum_{r=1}^3\epsilon^{\mu}_r(u(p))\epsilon^{\nu}_r(u(p)),
\nonumber \\
&&\eta_{AB}=\epsilon^{\mu}_A(u(p))\eta_{\mu\nu}\epsilon^{\nu}_B(u(p)),\nonumber 
\\
&&p_{\alpha}{{\partial}\over {\partial p_{\alpha}} }\epsilon^{\mu}_A(u(p))=
p_{\alpha}{{\partial}\over {\partial p_{\alpha}} }\epsilon^A_{\mu}(u(p))
=0.
\label{a13}
\end{eqnarray}

The Wigner rotation corresponding to the Lorentz transformation $\Lambda$ is

\begin{eqnarray}
R^{\mu}{}_{\nu}(\Lambda ,p)&=&{[L(\stackrel{\circ}{p},p)\Lambda^{-1}L(\Lambda
p,\stackrel{\circ}{p})]}^{\mu}{}_{\nu}=\left(
\begin{array}{cc}
1 & 0 \\
0 & R^i{}_j(\Lambda ,p)
\end{array}  
\right) ,\nonumber \\
{} && {}\nonumber \\
R^i{}_j(\Lambda ,p)&=&{(\Lambda^{-1})}^i{}_j-{ {(\Lambda^{-1})^i{}_op_{\beta}
(\Lambda^{-1})^{\beta}{}_j}\over {p_{\rho}(\Lambda^{-1})^{\rho}{}_o+\eta 
\sqrt{p^2}} }-\nonumber \\
&-&{{p^i}\over {p^o+\eta \sqrt{p^2}} }[(\Lambda^{-1})^o{}_j- { {((\Lambda^{-1})^o
{}_o-1)p_{\beta}(\Lambda^{-1})^{\beta}{}_j}\over {p_{\rho}(\Lambda^{-1})^{\rho}
{}_o+\eta \sqrt{p^2}} }].
\label{a14}
\end{eqnarray}

The polarization vectors transform under the 
Poincar\'e transformations $(a,\Lambda )$ in the following way

\begin{equation}
\epsilon^{\mu}_r(u(\Lambda p))=(R^{-1})_r{}^s\, \Lambda^{\mu}{}_{\nu}\, 
\epsilon^{\nu}_s(u(p)).
\label{a15}
\end{equation}

\vfill\eject

\section{Lagrangian for Dirac fields on Spacelike Hypersurfaces.}

In tetrad gravity, given a 3+1 splitting of a globally hyperbolic spacetime
$M^4$ with metric ${}^4g_{\mu\nu}$ [$\mu$ are world indices, $(\mu )$ are flat
rectangular Minkowski indices], one writes the following action for Dirac
fields $\tilde \psi (\tau ,\vec \sigma ) = \psi (z(\tau ,\vec \sigma ))$
(see Refs.\cite{naka,wei})

\begin{eqnarray}
S&=& \int d\tau d^3\sigma N(\tau ,\vec \sigma ) \sqrt{\gamma (\tau ,\vec 
\sigma )} \Big( {i\over 2} {\tilde \psi}^{\dagger} \gamma^{(o)} (\gamma^{(\mu )}
\, {}^4E^{\check A}_{(\mu )} {\vec D}_{\check A}-{\buildrel \leftarrow \over
D}_{\check A}\, {}^4E^{\check A}_{(\mu )} \gamma^{(\mu )})\tilde \psi -
\nonumber \\
&-&m{\tilde \psi}^{\dagger} \gamma^{(o)} \tilde \psi \Big) (\tau ,\vec \sigma )=
\int d\tau d^3\sigma {\cal L}(\tau ,\vec \sigma ),
\label{b1}
\end{eqnarray}

\noindent where

i) ${}^4E^{\check A}_{(\mu )}(\tau ,\vec \sigma )=b^{\check A}_{\mu}(\tau 
,\vec \sigma )\, {}^4E^{\mu}_{(\mu )}(\tau ,\vec \sigma )$ [$b^{\check A}_{\mu}
(\tau ,\vec \sigma )=\partial \sigma^{\check A}(z)/\partial z^{\mu} (\tau 
,\vec \sigma )$] are cotetrads in coordinates $\sigma^{\check A}=(\tau 
;\vec \sigma )$ adapted to the spacelike hypersurfaces $\Sigma_{\tau}$ leaves
of a foliation giving a 3+1 splitting of $M^4$
[${}^4E^{\check A}_{(\mu )}\, {}^4\eta^{(\mu )(\nu )}\, {}^4E^{\check B}
_{(\nu )}={}^4g^{\check A\check B}$ (${}^4\eta^{(\mu )(\nu )}$ is the flat
Minkowski inverse metric) with ${}^4g^{\check A\check B}(\tau ,\vec \sigma )$
inverse of the metric ${}^4g_{\check A\check B}=b^{\mu}_{\check A}b^{\nu}
_{\check B}\, {}^4g_{\mu\nu}={}^4E^{(\mu )}_{\check A}\, {}^4\eta_{(\mu )(\nu )}
\, {}^4E^{(\nu )}_{\check B}$, where ${}^4E^{(\mu )}_{\check A}(\tau ,\vec 
\sigma )=b^{\mu}_{\check A}(\tau ,\vec \sigma )\, {}^4E^{(\mu )}_{\mu}(\tau 
,\vec \sigma )$ are tetrads, ${}^4E^{(\mu )}_{\check A}\, {}^4E^{\check A}
_{(\nu )}=\delta^{(\mu )}_{(\nu )}$, ${}^4E^{(\mu )}_{\check A}\, {}^4E
^{\check B}_{(\mu )}=\delta^{\check B}_{\check A}$];

ii) $\gamma^{(\mu )}$ are flat Dirac matrices: $[\gamma^{(\mu )},\gamma
^{(\nu )}]_{+}=\gamma^{(\mu )}\gamma^{(\nu )}+\gamma^{(\nu )}\gamma^{(\mu )}=
2\, {}^4\eta^{(\mu )(\nu )}$, $\sigma^{(\mu )(\nu )}={i\over 2} [\gamma
^{(\mu )},\gamma^{(\nu )}]$;

iii) ${\vec D}_{\check A}=\partial_{\check A}-{i\over 4}\, {}^4\omega
_{\check A(\mu )(\nu )} \sigma^{(\mu )(\nu )}$ is the spinor covariant 
derivative ($\partial_{\check A}=\partial /\partial \sigma^{\check A}$) and 
${}^4\omega_{\check A(\mu )(\nu )}$ is the 4-spin connection [see Ref.
\cite{naka} for its expression in terms of cotetrads];

iv) $N(\tau ,\vec \sigma )$ is the lapse function and $\gamma (\tau ,\vec 
\sigma )=det\, {}^3g_{\check r\check s}(\tau ,\vec \sigma )$ [${}^3g_{\check 
r\check s}=-{}^4g_{\check r\check s}$].

See Ref.\cite{geh} for the 3+1 splitting of the 4-spin connection in terms of
the 3-spin connection [function of cotriads ${}^3e^{\check r}_i$ on $\Sigma
_{\tau}$], and lapse and shift functions. In the Hamiltonian version of tetrad
gravity cotriads, lapse and shift functions are the independent variables.
In this case the tetrads ${}^4E^{(\mu )}_{\mu}(\tau ,\vec \sigma )$ correspond 
to a $\Sigma_{\tau}$-adapted anholonomic basis of coordinates in the spacetime 
$M^4$.

When we consider a 3+1 splitting of Minkowski spacetime with a family of 
spacelike hypersurfaces $\Sigma_{\tau}$ the tetrads ${}^4E^{(\mu )}_{\check A}
(\tau ,\vec \sigma )$ and cotetrads ${}^4E^{\check A}_{(\mu )}(\tau ,\vec 
\sigma )$ go into the flat tetrads $z^{(\mu )}_{\check A}(\tau ,\vec \sigma )=
\partial z^{(\mu )}(\tau ,\vec \sigma )/\partial \sigma^{\check A}$ and 
cotetrads $z^{\check A}_{(\mu )}(\tau ,\vec \sigma )=\partial \sigma^{\check A}
(z)/\partial z^{(\mu )} (\tau ,\vec \sigma )$ respectively corresponding to a 
holonomic basis for Minkowski spacetime. Since in this paper we consider only 
Minkowski spacetime, from now on we shall drop the distinction betwenn flat 
and world indices, $z^{(\mu )}(\tau ,\vec \sigma )=z^{\mu}(\tau ,\vec \sigma )$.

Since in the holonomic basis for
Minkowski spacetime the 4-spin connection vanishes, ${}^4\omega
_{\check A \mu\nu}(\tau ,\vec \sigma )=\eta_{\mu\alpha}{^4\omega_{\bar{A}\nu}
^\alpha} (\tau,\vec{\sigma})=-^4\omega_{\bar{A}\nu\mu}(\tau, \vec{\sigma})
=0$, 

\begin{eqnarray}
{}^4\Gamma_{\check A\check C}^{\check B}(\tau, \vec{\sigma})&=&
\frac{1}{2}\, {}^4g^{\check B\check D}(\tau, \vec{\sigma})\left[
\partial_{\check A}\, {}^4g_{\check C\check D}(\tau,\vec{\sigma})+
\partial_{\check C}\,  {}^4g_{\check A\check D}(\tau,\vec{\sigma}) -
\partial_{\check D}\,  {}^4g_{\check A\check C}(\tau,\vec{\sigma})\right] =0,
\nonumber \\
&\Downarrow& \nonumber \\
{}^4\omega_{\check A \nu}^\alpha(\tau, \vec{\sigma})
&=&z_{\check B}^\alpha(\tau, \vec{\sigma})
\left[\partial_{\check A}z_\nu^{\check B}(\tau,
\vec{\sigma})+{}^4\Gamma_{\check A\check C}^{\check B}(\tau,
\vec{\sigma})z_\nu^{\check C}(\tau, \vec{\sigma})\right] =0,
\label{b2}
\end{eqnarray}

\noindent the action for Dirac fields becomes

\begin{equation}
S= \int d\tau d^3\sigma N(\tau ,\vec \sigma ) \sqrt{\gamma (\tau ,\vec 
\sigma )} [{i\over 2} {\tilde \psi}^{\dagger} \gamma^o (\gamma^{\mu} 
z^{\check A}_{\mu} {\vec \partial}_{\check A}-{\buildrel \leftarrow \over 
\partial}_{\check A}\, z^{\check A}_{\mu} \gamma^{\mu})\tilde \psi 
-m {\tilde \psi}^{\dagger} \gamma^o\tilde \psi ](\tau ,\vec \sigma ).
\label{b3}
\end{equation}

The normal $l^{\mu}(\tau ,\vec \sigma )=[{1\over {\sqrt{\gamma}}} \epsilon^{\mu}
{}_{\alpha\beta\gamma}z^{\alpha}_{\check 1}z^{\beta}_{\check 2}z^{\gamma}
_{\check 3}](\tau ,\vec \sigma )$ to $\Sigma_{\tau}$ is timelike, so that if 
${\buildrel \circ \over l}^{\mu}=(1;\vec 0)$, one has $l^{\mu}(\tau ,\vec 
\sigma )=L^{\mu}{}_{\nu}(l(\tau ,\vec \sigma ),{\buildrel \circ \over l})
{\buildrel \circ \over l}^{\nu}$ by using the corresponding Wigner boost. One 
can then define new flat (nonholonomic) tetrads ${\buildrel \circ \over z}
^{\mu}_{\check A}(\tau ,\vec \sigma )$ through $z^{\mu}_{\check A}(\tau ,\vec 
\sigma )=L^{\mu}{}_{\nu}(l(\tau ,\vec \sigma ),{\buildrel \circ \over l})
{\buildrel \circ \over z}^{\nu}_{\check A}(\tau ,\vec \sigma )$ and the 
corresponding cotetrads through $z^{\check A}_{\mu}(\tau ,\vec \sigma )=
{\buildrel \circ \over z}^{\check A}_{\nu}(\tau ,\vec \sigma ) L^{\nu}{}_{\mu}
(l(\tau ,\vec \sigma ),{\buildrel \circ \over l})$. One finds

\begin{eqnarray}
{\buildrel o \over {z_{\tau}^\mu}}(\tau, \vec{\sigma})
&=&\left(N(\tau, \vec{\sigma});
N^{\breve{r}}(\tau, \vec{\sigma})~^3e_{\breve{r}}^i(\tau, \vec{\sigma})\right),
\nonumber \\
{\buildrel o \over {z_{\breve{r}}}^\mu}(\tau, \vec{\sigma})&=&
\left (0;^3e_{\breve{r}}^i(\tau, \vec{\sigma})\equiv z_{\breve{r}}^i(\tau,
\vec{\sigma})-\frac{l^i(\tau, \vec{\sigma})}{1+l^0(\tau, \vec{\sigma})}
z^{\tau}_{\breve{r}}(\tau, \vec{\sigma})\right),\nonumber \\
&&{}\nonumber \\
{\buildrel o \over {z^\tau_\mu}}(\tau, \vec{\sigma})
&=&\left(\frac{1}{N(\tau, \vec{\sigma})}
;\vec{0}\right) ,\nonumber \\
{\buildrel o \over {z_\mu^{\breve{r}}}}(\tau, \vec{\sigma})&=&
\left(-\frac{N^{\breve{r}}}{N} (\tau, \vec{\sigma});
 ^3e_i^{\breve{r}}(\tau, \vec{\sigma})=-\gamma^{\check r\check s}(\tau ,\vec 
\sigma )\, {}^3e_{\check r i}(\tau ,\vec \sigma ) \right) ,
\label{b4}
\end{eqnarray}

\noindent where now $N=\sqrt{g/\gamma}$, $N^{\check r}=g_{\tau \check s}
\gamma^{\check s\check r}$ (see Appendix A).

The metric becomes

\begin{eqnarray}
&&g_{\breve{A}\breve{B}}(\tau, \vec{\sigma})= z_{\breve {A}}^\mu
(\tau, \vec{\sigma})\eta_{\mu\nu} z^\nu_{\breve {B}}(\tau, \vec{\sigma})
={\buildrel o \over {z_{\breve{A}}^\mu}} (\tau, \vec{\sigma}) \eta_{\mu \nu}
{\buildrel o \over {z_{\breve{B}}^\nu}}(\tau, \vec{\sigma}),\nonumber \\
&&{}^3g_{\breve {r}\breve{s}}(\tau, \vec{\sigma})=
{\buildrel o \over {z^\mu_{\breve{r}}}}(\tau, \vec{\sigma})\eta_{\mu\nu}
{\buildrel o \over {z_{\breve{s}}^\nu}}(\tau, \vec{\sigma})
=-^3e^i_{\breve{r}}(\tau, \vec{\sigma})
^3e_{\breve{s}}^i(\tau, \vec{\sigma}),\nonumber \\
&&{}\nonumber \\
&&g^{\breve{A}\breve{B}}(\tau, \vec{\sigma})=z^{\breve{A}}_\mu(\tau,
\vec{\sigma})\eta^{\mu\nu}z_\nu^{\breve{B}}(\tau, \vec{\sigma})=
{\buildrel o \over {z_\mu^{\breve{A}}}}(\tau, \vec{\sigma}) \eta^{\mu\nu}
{\buildrel o \over {z^{\breve{B}}_\nu}}(\tau, \vec{\sigma}).
\label{b5}
\end{eqnarray}

In this case cotriads ${}^3e^{\check r}_i(\tau, \vec{\sigma})$, triads
${}^3e^i_{\check r}(\tau, \vec{\sigma})$\hfill\break
[$^3e^i_{\breve{r}}(\tau, \vec{\sigma})^3e_i^{\breve{s}}(\tau,
\vec{\sigma})=\delta_{\breve{r}}^{\breve{s}},~~~~~^3e_{\breve{r}}^i(\tau,
\vec{\sigma})^3e_j^{\breve{r}}(\tau, \vec{\sigma})=\delta^i_j$],
lapse $N(\tau, \vec{\sigma})$ and shifts $N^{\check r}(\tau, \vec{\sigma})$ 
[whose expression is given in Appendix A] are all functionals of the
independent variables $z^{\mu}(\tau, \vec{\sigma})$ [which do not exist in
general relativity, which has no global holonomic basis], the coordinates 
describing the embedding of 3-surfaces as spacelike hypersurfaces $\Sigma
_{\tau}$ in Minkowski spacetime.

Let us now consider point dependent Lorentz transformations

\begin{eqnarray}
\psi(\tau, \vec{\sigma})\rightarrow\psi^\prime(\tau,
\vec{\sigma})&=&S(\Lambda(\tau, \vec{\sigma}))\psi(\tau, \vec{\sigma}),
\nonumber \\
\bar{\psi}(\tau, \vec{\sigma})\rightarrow{\bar{\psi}}^\prime(\tau,
\vec{\sigma})&=&
\bar{\psi}(\tau, \vec{\sigma})S^{-1}(\Lambda(\tau, \vec{\sigma})).
\label{b6}
\end{eqnarray}

\noindent and their action on  the 4-spin connection

\begin{equation}
{}^4\omega_{\check{A}}(\tau, \vec{\sigma})
{\buildrel o \over {\Lambda(\tau, \vec{\sigma})}}
\longrightarrow  S(\Lambda(\tau, \vec{\sigma}))\, {}^4\omega_{\check{A}}(\tau,
\vec{\sigma})S^{-1}(\Lambda(\tau, \vec{\sigma}))-\partial_{\check{A}}
S(\Lambda(\tau, \vec{\sigma}))S^{-1}(\Lambda(\tau, \vec{\sigma})).
\label{b7}
\end{equation}

\noindent for $\Lambda (\tau, \vec{\sigma})=L(l(\tau ,\vec \sigma ),{\buildrel 
\circ \over l})$ [see Appendix C for the associated Lorentz transformation].
Then, by using

\begin{eqnarray}
{\buildrel \circ \over {\tilde \psi}}(\tau ,\vec \sigma )&=&S^{-1}(L)\tilde \psi
,\nonumber \\
D_{\check A}&=&\partial_{\check A}-{}^4{\buildrel \circ \over \omega}_{\check 
A},\nonumber \\
0={}^4\omega_{\check A}&=&S(L)\, {}^4{\buildrel \circ \over \omega}_{\check A}
 S^{-1}(L)+\partial_{\check A}S(L)\, S^{-1}(L),\nonumber \\
&&\Rightarrow \, {}^4{\buildrel \circ \over \omega}_{\check A}=-S^{-1}(L) 
\partial_{\check A}S(L),\nonumber \\
\gamma^{\mu}z^{\check A}_{\mu}\partial_{\check A}\tilde \psi &=&S(L)\gamma
^{\mu}{\buildrel \circ \over z}^{\check A}_{\mu}{\buildrel \circ \over D}
_{\check A}{\buildrel \circ \over {\tilde \psi}}.
\label{b8}
\end{eqnarray}

\noindent we get the following form for the action for Dirac fields

\begin{equation}
S=\int d\tau d^3\sigma N(\tau ,\vec \sigma )\sqrt{\gamma(\tau ,\vec \sigma )}
[{i\over 2}{\buildrel \circ \over {\bar {\tilde \psi}}}
(\gamma^{\mu}{\buildrel \circ \over z}^{\check A}_{\mu}{\buildrel \circ \over
{\vec D}}_{\check A}-{\buildrel \circ \over {\buildrel \leftarrow \over D}}
_{\check A}{\buildrel \circ \over z}^{\check A}_{\mu}\gamma^{\mu}) {\buildrel
\circ \over {\tilde \psi}}
-m{\buildrel \circ \over {\bar {\tilde \psi}}}{\buildrel \circ \over
{\tilde \psi}}](\tau ,\vec \sigma ),
\label{b9}
\end{equation}

An analogous action could be written in tetrad gravity by using the expression
of the 4-spin connection in terms of cotriads, lapse and shifts 
and this is the form of action to be used for defining the Hamiltonian 
description of fermion fields in tetrad gravity
[one could also use Eq.(\ref{b1}) with the 4-spin connection expressed in
terms of cotriads, lapse and shifts]. 

In Minkowski spacetime it is more convenient to use Eq.(\ref{b3})
[namely Eq.(\ref{b1}) in a holonomic basis] rather than either Eq.(\ref{b9})
or Eq.(\ref{b1}) in other bases. 

For instance in pseudoclassical electrodynamics in 
an arbitrary basis, Eq.(\ref{b1}) becomes [for the sake of simplicity we write 
$\psi$ for $\tilde \psi$ and $\bar \psi$ for ${\tilde \psi}^{\dagger}\gamma^o$]

\begin{eqnarray}
{\cal{L}(\tau, \vec{\sigma})}
&=&N(\tau, \vec{\sigma})\sqrt{\gamma(\tau, \vec{\sigma})}
\{
\frac{i}{2}{\bar \psi}(\tau, \vec{\sigma})\gamma^\mu 
z^{\check A}_\mu(\tau, \vec{\sigma})(
{\vec D}_{\check A}-{\buildrel \leftarrow \over 
D}_{\check A})\psi(\tau, \vec{\sigma}) +\nonumber \\
 &+&e A_{\check A}(\tau, \vec{\sigma}){\bar \psi}(\tau,
\vec{\sigma})\gamma^\mu z_\mu^{\check A}(\tau,
\vec{\sigma})\psi(\tau, \vec{\sigma})-m
{\bar \psi}(\tau,\vec{\sigma})\psi (\tau, \vec{\sigma})+\nonumber \\
&-&\frac{1}{4}g^{\check A\check C}(\tau,\vec{\sigma})g^{\check B\check D}(\tau, 
\vec{\sigma})F_{\check A\check B}(\tau, \vec{\sigma}) F_{\check C\check D}
(\tau, \vec{\sigma}) \} =\nonumber \\
&=&N(\tau, \vec{\sigma})\sqrt{\gamma(\tau, \vec{\sigma})}\{
\frac{i}{2}\bar{\psi}(\tau, \vec{\sigma})\gamma^\mu z^{\check A}_\mu(\tau,
\vec{\sigma})\partial_{\check A}\psi(\tau, \vec{\sigma})-\frac{i}{2}
\partial_{\check A}\bar{\psi}(\tau, \vec{\sigma})\gamma^\mu z^{\check A}_\mu
(\tau,\vec{\sigma})\psi(\tau, \vec{\sigma})+\nonumber \\
&+&\frac{1}{8}\bar{\psi}(\tau, \vec{\sigma}) {}^4\omega_{\check A
\alpha\beta}(\tau, \vec{\sigma})\left[ \gamma^\mu, \sigma^{\alpha\beta}
\right]_+
z^{\check A}_\mu(\tau, \vec{\sigma})\psi(\tau, \vec{\sigma})
-m\bar{\psi}(\tau, \vec{\sigma})\psi(\tau, \vec{\sigma})\} +\nonumber \\
 &+&e A_{\check A}(\tau, \vec{\sigma}){\bar \psi}(\tau,
\vec{\sigma})\gamma^\mu z_\mu^{\check A}(\tau,
\vec{\sigma})\psi(\tau, \vec{\sigma})-m
{\bar \psi}(\tau,\vec{\sigma})\psi (\tau, \vec{\sigma})+\nonumber \\
&-&\frac{1}{4}g^{\check A\check C}(\tau,\vec{\sigma})g^{\check B\check D}(\tau, 
\vec{\sigma})F_{\check A\check B}(\tau, \vec{\sigma}) F_{\check C\check D}
(\tau, \vec{\sigma}) \} ,
\label{b10}
\end{eqnarray}

\noindent to be compared with Eq.(\ref{b3}) in the holonomic basis   for the
Dirac subsystem.

In the nonholonomic basis (\ref{b4}) and without the electromagnetic
field for the sake of simplicity, Eq.(\ref{b1}) becomes

\begin{eqnarray}
{\cal{L}}(\tau, \vec{\sigma})&=&
N(\tau, \vec{\sigma})\sqrt{\gamma(\tau, \vec{\sigma})}
\{
\frac{i}{2}{\buildrel \circ \over {\bar \psi}}(\tau, \vec{\sigma})\gamma^\mu 
{\buildrel \circ \over z}^{\check A}_\mu(\tau, \vec{\sigma})({\buildrel \circ
\over {\vec D}}_{\check A}-{\buildrel \circ \over {\buildrel \leftarrow \over 
D}}_{\check A}){\buildrel \circ \over \psi}(\tau, \vec{\sigma}) +\nonumber \\
 &+&e A_{\check A}(\tau, \vec{\sigma}){\buildrel \circ \over {\bar \psi}}(\tau,
\vec{\sigma})\gamma^\mu {\buildrel \circ \over z}_\mu^{\check A}(\tau,
\vec{\sigma}){\buildrel \circ \over \psi}(\tau, \vec{\sigma})-m{\buildrel
\circ \over {\bar \psi}}(\tau,\vec{\sigma}){\buildrel \circ \over \psi}(\tau, 
\vec{\sigma})+\nonumber \\
&-&\frac{1}{4}g^{\check A\check C}(\tau,\vec{\sigma})g^{\check B\check D}(\tau, 
\vec{\sigma})F_{\check A\check B}(\tau, \vec{\sigma}) F_{\check C\check D}
(\tau, \vec{\sigma}) \} =\nonumber \\
&=&\sqrt{\gamma(\tau, \vec{\sigma})}\{
\frac{i}{2}\left( {\buildrel \circ \over {\bar \psi}}(\tau, \vec{\sigma})
\gamma^0\partial_\tau {\buildrel \circ \over \psi}
(\tau,\vec{\sigma})-\partial_\tau {\buildrel \circ \over {\bar \psi}}(\tau,
\vec{\sigma})\gamma^0{\buildrel \circ \over \psi}(\tau, \vec{\sigma})\right)+
\nonumber \\
&-&\frac{i}{2}N^{\check r}(\tau, \vec{\sigma})
\left( {\buildrel \circ \over {\bar \psi}}(\tau, \vec{\sigma})
\gamma^0\partial_{\check r}{\buildrel \circ \over \psi}(\tau,
\vec{\sigma})-\partial_{\bar{r}}\bar{\psi}(\tau, \vec{\sigma})
\gamma^0 \psi(\tau,\vec{\sigma})\right)+\nonumber \\
&-&\frac{i}{8}{\cal{F}}_{ij}(\tau, \vec{\sigma}){\buildrel \circ \over {\bar 
\psi}}(\tau,\vec{\sigma})\gamma^0\gamma^i\gamma^j{\buildrel \circ \over \psi}
(\tau, \vec{\sigma})-\frac{i}{2}\left(Tr^3{{K}}(\tau, \vec{\sigma})\right)
{\buildrel \circ \over {\bar \psi}}(\tau,\vec{\sigma})\gamma^0
{\buildrel \circ \over \psi}(\tau, \vec{\sigma})\}+\nonumber \\
&+&N(\tau, \vec{\sigma})\sqrt{\gamma(\tau, \vec{\sigma})}
\{\frac{i}{2}{\buildrel \circ \over {\bar \psi}}(\tau,
\vec{\sigma})^3e^{\check r}_i(\tau, \vec{\sigma})\gamma^i
\left(\partial_ {\check r}{\buildrel \circ \over \psi}
(\tau, \vec{\sigma})+\frac{1}{4}{^4\omega_{\check rjk}}(\tau,
\vec{\sigma})\gamma^j\gamma^k{\buildrel \circ \over \psi}(\tau, \vec{\sigma})
\right)+\nonumber \\
&-&\frac{i}{2}\left(\partial_{\check r}{\buildrel \circ \over {\bar \psi}}
(\tau, \vec{\sigma})-\frac{1}{4}
{^4\omega_{\check rjk}}(\tau, \vec{\sigma}){\buildrel \circ \over {\bar \psi}}
(\tau,\vec{\sigma})\gamma^k\gamma^j\right)~^3e_i^{\check r}(\tau,
\vec{\sigma})\gamma^i{\buildrel \circ \over \psi}(\tau, \vec{\sigma})-m
{\buildrel \circ \over {\bar \psi}}(\tau,\vec{\sigma}){\buildrel \circ \over 
\psi}(\tau, \vec{\sigma})\} ,\nonumber \\
&&{}\nonumber \\
\label{b11}
\end{eqnarray}

\noindent where [${}^3K_{ij}$ is the extrinsic curvature of $\Sigma_{\tau}$ 
embedded in Minkowski spacetime]

\begin{eqnarray}
{\cal{F}}_{ij}(\tau, \vec{\sigma})
&\equiv&\left({^3e_{i{\check r}}}(\tau,
\vec{\sigma})~{^3e_j^{\check s}} (\tau,
\vec{\sigma})+3~{^3e_{j{\check r}}}(\tau,
\vec{\sigma})~{^3e_i^{\check s}} (\tau, \vec{\sigma})
\right)\partial_{\check s}N^{\check r}(\tau, \vec{\sigma}) +\nonumber \\
&-&N^{\check r}(\tau, \vec{\sigma})\left({^3e_{i\check s}}(\tau,
\vec{\sigma})
\partial_{\check r}{}^3e_j^{\check s}(\tau, \vec{\sigma})  +3~{^3e_{j{\check 
s}}}(\tau, \vec{\sigma})  \partial_{\check r}\, {}^3e_i^{\check s}(\tau, 
\vec{\sigma})\right)+\nonumber \\
&+&{^3e_{i\check r}}(\tau, \vec{\sigma})
\partial_\tau~{^3e_j^{\check r}} (\tau,
\vec{\sigma})+3~{^3e_{j\check r}}(\tau, \vec{\sigma})
\partial_\tau~{^3e_i^{\check r}}(\tau, \vec{\sigma}) ,\nonumber \\
^3{{K}}_{\check r\check s}(\tau, \vec{\sigma})
&=&\frac{1}{2N(\tau, \vec{\sigma})  }\left[
\partial_{\check s}N_{\check r}(\tau, \vec{\sigma})  -\partial_{\check r}
N_{\check s}(\tau, \vec{\sigma})  -\partial_\tau\,  {}^3g_{\check r\check s}
(\tau,\vec{\sigma})\right] ,\nonumber \\
Tr^3{{K}}(\tau, \vec{\sigma}) &\equiv&
-{^3e^i_{\check r}}(\tau, \vec{\sigma})\,  ^3e_i^{\check s}(\tau,
\vec{\sigma})  \partial_{\check s}N^{\check r}(\tau, \vec{\sigma})  +
N^{\check r}(\tau, \vec{\sigma})\,  ^3e_{\check s}^i(\tau,
\vec{\sigma})
\partial_{\check r}\, {^3e_i^{\check s}}(\tau, \vec{\sigma})  +\nonumber \\
&-&{^3e_{\check r}}^i(\tau, \vec{\sigma})  \partial_\tau \,
{^3e_i^{\check r}} (\tau, \vec{\sigma}).
\label{b12}
\end{eqnarray}

\vfill\eject

\section{Transformation Properties of the Dirac Field under Wigner Boosts.}

We can give an exponential form\cite{longhi} of the Wigner boost of Eq.
(\ref{a9})

\begin{eqnarray}
L^\mu{}_{\nu} (p, \stackrel{o}p) &=&
\exp \big[ \omega (p) I (p) \big]^\mu{}_{\nu} = \nonumber \\
&=& \big[ \cosh\big(\omega (p) I (p)\big)
+ \sinh \big(\omega (p) I (p)\big) \big]^\mu{}_{\nu} = \nonumber \\
&=& \big[ {I} - I^2(p) + I^2(p)
\cosh \omega (p)
+ I(p) \sinh \omega (p) \big]^\mu{}_{\nu}, \nonumber \\
 L^\mu{}_{\nu}(\stackrel{o}p, p) &=&
\exp \big[ - \omega (p) I (p) \big]^\mu{}_{\nu},
\label{c1}
\end{eqnarray}

\noindent where

\begin{eqnarray}
\cosh \omega(p) &=& \frac{\eta p_0}{\sqrt{p^2}}, \quad\quad
\sinh \omega(p) = \eta \frac{|\vec{p}|}{\sqrt{p^2}}, \nonumber \\
I(p) &\equiv & \parallel I(p)^\mu_{.\nu} \parallel =
\pmatrix{0& -\frac{p_j}{|\vec{p}|}\cr \frac{p^i}{|\vec{p}|}& 0},\nonumber \\
&&{}\nonumber \\
I_{\mu \nu}(p) &=& - I_{\nu \mu}(p),~~~~~I^3(p) = I(p)  .
\label{c2}
\end{eqnarray}

If we consider the generating function of the canonical transformation of
Eq.(\ref{IV1})

\begin{equation}
{\cal F}(p_s) = \frac{1}{2} \omega(p_s) I_{\mu \nu}(p_s) S^{\mu \nu},
\label{c3}
\end{equation}

\noindent with $S^{\mu\nu}$ the total spin of Eq.(\ref{III12}) [and not only
$S_s^{\mu\nu}$ as one would have for scalar fields], we get

\begin{eqnarray}
\stackrel{o}\psi (\tau , \vec{\sigma}) &=&
\exp \{ {\cal F} , .\}^\ast_D \psi (\tau, \vec{\sigma}) \equiv \nonumber \\
&=& \psi (\tau , \vec{\sigma}) + \{ {\cal F} ,
\psi (\tau , \vec{\sigma}) \}^\ast_D
+\frac{1}{2} \{ {\cal F}, \{ {\cal F}, \psi (\tau , \vec{\sigma}) \}^\ast_D
\}^\ast_D + ... = \nonumber \\
&=& \psi (\tau , \vec{\sigma}) + \frac i4 \omega (p_s) I_{\mu \nu} (p_s)
\sigma^{\mu \nu} \psi (\tau , \vec{\sigma}) + ... = \nonumber \\
&=& \exp\big[ \frac i4 \omega (p_s) I_{\mu \nu} (p_s)
\sigma^{\mu \nu} \big]\psi (\tau , \vec{\sigma}).
\label{c4}
\end{eqnarray}

This shows that this canonical transformation implements on the Dirac fields 
the action of Wigner boosts realized by using the standard representation of 
Lorentz transformations in terms of Dirac matrices\cite{itz}

\begin{equation}
S(L(\stackrel{o}p_s, p_s)) = \exp \big[ \frac i4 \omega (p_s) I_{\mu \nu} (p_s)
\sigma^{\mu \nu} \big]    .
\label{c5}
\end{equation}

Since we have from Section IV

\begin{eqnarray}
\{ \tilde{x}^\mu_s , \stackrel{o}\psi (\tau , \vec{\sigma}) \}^\ast_D &=&
\{ \tilde{x}^\mu_s , S(L(\stackrel{o}p_s, p_s))
\psi (\tau , \vec{\sigma}) \}^\ast_D  = \nonumber \\
&=& - \frac{\partial S(L(\stackrel{o}p_s, p_s))}{\partial {p_s}_\mu}
S^{-1} (L(\stackrel{o}p_s, p_s)) \stackrel{o}\psi (\tau , \vec{\sigma}) + 
\nonumber \\
&-& \frac i4 \epsilon^A_\nu (u(p_s)) \eta_{AB}
\frac{\partial \epsilon^B_\rho (u(p_s))} {\partial {p_s}_\mu}
S(L(\stackrel{o}p_s, p_s)) \sigma^{\nu \rho}
S^{-1}(L(\stackrel{o}p_s, p_s))
 \stackrel{o}\psi (\tau , \vec{\sigma}) = \nonumber \\
&=& \Big[ - \frac{\partial S(L(\stackrel{o}p_s, p_s))}{\partial {p_s}_\mu}
S^{-1} (L(\stackrel{o}p_s, p_s)) + \nonumber \\
&-& \frac i4 \eta_{\sigma B}
\frac{\partial \epsilon^B_\rho (u(p_s))} {\partial {p_s}_\mu}
L^\rho{}_{\eta} (p_s, \stackrel{o}p_s) \sigma^{\sigma \eta} \Big]
 \stackrel{o}\psi (\tau , \vec{\sigma}) ,
\label{c6}
\end{eqnarray}

\noindent to verify $\{ \tilde{x}^\mu_s , \stackrel{o}\psi (\tau , 
\vec{\sigma}) \}^\ast_D = 0$, see Eq.(\ref{IV4}), we need the evaluation of
$\frac{\partial S(L(\stackrel{o}p_s, p_s))}{\partial {p_s}_\mu}$ .

In Ref.\cite{wilcox}, there is the following formula

\begin{equation}
\frac{\partial e^{B(\lambda)}}{\partial \lambda} =
\int^{1}_{0} dx e^{xB(\lambda)} \frac{\partial B(\lambda)}{\partial \lambda}
e^{-xB(\lambda)} e^{B(\lambda)},
\label{c7}
\end{equation}

\noindent giving the derivative with respect to a continous parameter $\lambda$
of the exponential of an operator $B(\lambda )$. If we put

\begin{eqnarray}
A(p_s) &\equiv& \frac i4 \omega (p_s) I_{\mu \nu} (p_s) \sigma^{\mu \nu} =
- \frac i2 \frac{\omega(p_s)}{|\vec{p_s}|} {p_s}_i \sigma^{0i},\nonumber \\
&&S(L(\stackrel{o}p_s, p_s)) = e^{A(p_s)},
\label{c8}
\end{eqnarray}

\noindent and if we suppose $p^\mu_s = p_s^\mu(\lambda)$, we have

\begin{eqnarray}
a)~\frac{\partial e^{A(p_s(\lambda))}}{\partial \lambda} &=&
\frac{\partial {p_s}_\mu(\lambda)}{\partial \lambda}
\frac{\partial e^{A(p_s(\lambda))}}{\partial {p_s}_\mu}, \nonumber \\
b)~\frac{\partial e^{A(p_s(\lambda))}}{\partial \lambda} &=&
\int^{1}_{0} dx e^{xA(p_s(\lambda))}
\frac{\partial A(p_s(\lambda))}{\partial \lambda}
e^{-xA(p_s(\lambda))} e^{A(p_s(\lambda))}  = \nonumber \\
&=& \frac{\partial {p_s}_\mu(\lambda)}{\partial \lambda}
\int^{1}_{0} dx e^{xA(p_s(\lambda))}
\frac{\partial A(p_s(\lambda))}{\partial {p_s}_\mu}
e^{-xA(p_s(\lambda))} e^{A(p_s(\lambda))}.
\label{c9}
\end{eqnarray}

This implies

\begin{equation}
\frac{\partial S(L(\stackrel{o}p_s, p_s))}{\partial {p_s}_\mu} =
\frac{\partial e^{A(p_s)}}{\partial {p_s}_\mu} =
\int^{1}_{0} dx e^{xA(p_s)}
\frac{\partial A(p_s)}{\partial {p_s}_\mu}
e^{-xA(p_s)} e^{A(p_s)} .
\label{c10}
\end{equation}

Following Ref.\cite{wilcox}, the solution of this equation is

\begin{equation}
\frac{\partial e^{A(p_s)}}{\partial {p_s}_\mu} =
\sum_{n=0}^\infty \frac{\Big[ A^n(p_s),
\frac{\partial A(p_s)}{\partial {p_s}_\mu} \Big]}{\big( n + 1\big) !}
e^{A(p_s)},
\label{c11}
\end{equation}

\noindent where  $\big[ A^n , B \big]$ means

\begin{eqnarray}
\big[ A^0 , B \big] &=& B ,~~~~~\big[ A^1 , B \big] = AB - BA ,\nonumber \\
\big[ A^{n+1} , B \big] &=& \Big[ A, \big[ A^n , B \big] \Big] .
\label{c12}
\end{eqnarray}

From the following commutators of Dirac matrices

\begin{eqnarray}
\big[ \gamma^\mu , \gamma^\nu \big] &=&
-2 i \sigma^{\mu \nu}, \nonumber \\
\big[ \sigma^{\mu \nu} , \gamma^\rho \big] &=&
2 i \big( \gamma^\mu \eta^{\nu \rho} - \gamma^\nu \eta^{\mu \rho} \big) , 
\nonumber \\
\big[ \sigma^{\mu \nu} , \sigma^{\alpha \beta} \big] &=&
2 i C^{\mu \nu \alpha \beta}_{\gamma \delta} \sigma^{\gamma \delta},
\label{c13}
\end{eqnarray}

\noindent and using Eq.(\ref{c2}), we get

\begin{eqnarray}
\frac{\partial S(L(\stackrel{o}p_s, p_s))}{\partial {p_s}_\mu} &=&
\Big[ \frac i2 \frac {{p_s}_i \sigma^{0i}}{\epsilon_s^2 ({p_s}_0 +
\epsilon_s)} \Big(p_s^\mu +2 \epsilon_s \eta^\mu_0 \Big)
- \frac i2 \frac{\sigma^{0\mu}}{\epsilon_s} + \nonumber \\
&+& \frac i2 \frac {{p_s}_i \sigma^{i \mu}}{\epsilon_s ({p_s}_0 +
\epsilon_s)} \Big] S(L(\stackrel{o}p_s, p_s)) = \nonumber \\
&=& \Big[ - \frac i4 \eta_{\sigma B}
\frac{\partial \epsilon^B_\rho(u(p_s))}{\partial {p_s}_\mu}
L^\rho_{.\eta}(p_s, \stackrel{o}p_s) \sigma^{\sigma \eta} \Big]
S(L(\stackrel{o}p_s, p_s)).
\label{c14}
\end{eqnarray}

\vfill\eject


\begin{references}


\bibitem{lusa}L.Lusanna, Int.J.Mod.Phys. {\bf A10}, 3531 and 3675 (1995).
\bibitem{lv1}L.Lusanna and P.Valtancoli, ``Dirac's Observables for the Higgs
model: I) the Abelian Case", Int.J.Mod.Phys. {\bf A12}, 4769 (1997) 
(HEP-TH/9606078).
\bibitem{lv2}L.Lusanna and P.Valtancoli, ``Dirac's Observables for the Higgs
model: II) the non-Abelian SU(2) Case", Int.J.Mod.Phys. {\bf A12}, 4797 (1997)
(HEP-TH/9606079).
\bibitem{lv3}L.Lusanna and P.Valtancoli, ``Dirac's Observables for the 
SU(3)xSU(2)xU(1) Standard Model", to appear in Int.J.Mod.Phys. A
(HEP-TH/9707072).
\bibitem{dira}P.A.M.Dirac, Can.J.Phys. {\bf 33}, 650 (1955).
\bibitem{sha}S.Shanmugadhasan, J.Math.Phys. {\bf 14}, 677 (1973).
L.Lusanna, Int.J.Mod.Phys. {\bf A8}, 4193 (1993).
M.Chaichian, D.Louis Martinez and L.Lusanna, Ann.Phys.(N.Y.){\bf 232}, 40 
(1994). L.Lusanna, 
Phys.Rep. {\bf 185}, 1 (1990); Riv. Nuovo Cimento {\bf 14}, n.3, 1 (1991);
J.Math.Phys. {\bf 31}, 2126 (1990); J.Math.Phys. 
{\bf 31}, 428 (1990).
\bibitem{re}L.Lusanna, ``Solving Gauss' Laws and Searching Dirac Observables
for the Four Interactions", talk at the ``Second Conf. on Constrained Dynamics
and Quantum Gravity", S.Margherita Ligure 1996, eds. V.De Alfaro, J.E.Nelson,
G.Bandelloni, A.Blasi, M.Cavagli\`a and A.T.Filippov, Nucl.Phys. (Proc.Suppl.)
{\bf B57}, 13 (1997) (HEP-TH/9702114). ``Unified
Description and Canonical Reduction to Dirac's Observables of the Four
Interactions", talk at the Int.Workshop ``New non Perturbative Methods and
Quantization on the Light Cone', Les Houches School 1997, eds. P.Grang\'e,
H.C.Pauli, A.Neveu, S.Pinsky and E.Werner (Springer, Berlin, 1998)
 (HEP-TH/9705154). ``The Pseudoclassical Relativistic Quark Model in the 
Rest-Frame Wigner-Covariant Gauge", talk given at the Euroconference QCD 97, 
Montpellier 1997, ed.S.Narison, Nucl.Phys. (Proc.Suppl.) {\bf B64}, 306 (1998).
\bibitem{lus1}L.Lusanna, Int.J.Mod.Phys. {\bf A12}, 645 (1997).
\bibitem{lus2}D.Alba and L.Lusanna, ``The Lienard-Wiechert Potential of Charged
Scalar Particles and their Relation to Scalar Electrodynamics in the Rest-Frame
Instant Form", to appear in Int.J.Mod.Phys. A (HEP-TH/9705155).
\bibitem{dplr}L.Lusanna and S.Russo, ``Tetrad Gravity: I) A New Formulation;
II) Dirac's Observables ", in preparation.\hfill\break
R.De Pietri and L.Lusanna, ``Tetrad Gravity: III) Asymptotic Poincar\'e Charges,
Void Spacetimes and the Physical Hamiltonian", in preparation.
\bibitem{lus3}D.Alba and L.Lusanna, ``The Classical Relativistic Quark Model in
the Rest-Frame Wigner-Covariant Coulomb Gauge", to appear in Int.J.Mod.Phys. A
(HEP-TH/9705156).
\bibitem{wei}S.Weinberg, Gravitation and Cosmology (J.Wiley, New York, 1972).
\bibitem{naka}M.Nakahara, Geometry, Topology and Physics (IOP, Bristol, 1990).
\bibitem{geh}B.S.DeWitt and C.M.DeWitt, Phys.Rev. {\bf 87}, 116 (1952).
\hfill\break
T.W.B.Kibble, J.Math.Phys. {\bf 4}, 1433 (1963).\hfill\break
P.A.M.Dirac, in ``Recent Development in General relativity" (Pergamon,
1962).\hfill\break
M.Henneaux, Gen.Rel.Grav. {\bf 9}, 1031 (1978).\hfill\break
J.Geheniau and M.Henneaux, Gen.Rel.Grav. {\bf 8}, 611 (1977).\hfill\break
J.E.Nelson and C.Teitelboim, Phys.Lett. {\bf 69B}, 81 (1977); Ann.Phys.(N.Y.)
{\bf 116}, 86 (1978).\hfill\break
P.D.D'Eath and J.J.Halliwell, Phys.Rev. {\bf D35}, 1100 (1987).
\bibitem{big}F.Bigazzi and L.Lusanna, ``Spinning Particles on Spacelike 
Hypersurfaces and their Rest-Frame Description", Firnze Univ. preprint 1997.
\bibitem{cha}A.Chakrabarti, J.Math.Phys. {\bf 4}, 1215 (1963).
\bibitem{mate}G.Longhi and M.Materassi, ``A Canonical Realization of the BMS 
Algebra", to appear in J.Math.Phys. (HEP-TH/9803128).
G.Longhi and M.Materassi, ``Collective and Relative Variables for a 
Klein-Gordon Field", in preparation.
\bibitem{sp1}E.Ramos and J.Roca, Nucl.Phys. {\bf B436}, 529 (1995).
\bibitem{fv}H.Feshbach and F.Villars, Rev.Mod.Phys. {\bf 30}, 24 (1958).
\bibitem{fw}L.L.Foldy and S.A.Wouthuysen, Phys.Rev. {\bf 78}, 29 (1950).
\bibitem{g5}A.Barducci and L.Lusanna, Nuovo Cimento {\bf 77A}, 39 (1983).
\bibitem{kuchar}K.Kuchar, J.Math.Phys. {\bf 17}, 777, 792, 801 (1976); {\bf 18},
1589 (1977).
\bibitem{longhi}G.Longhi and L.Lusanna, Phys.Rev. {\bf D34}, 3707 (1986).
\bibitem{itz}C.Itzykson and J.B.Zuber, Quantum Field Theory (McGraw Hill,
New York, 1985).
\bibitem{wilcox}R.M.Wilcox, J.Math.Phys. {\bf 8}, 962 (1967).
\bibitem{cou}K.Johnson, Ann.Phys.(N.Y.) {\bf 10}, 536 (1960).
R.Hagen, Phys.Rev. {\bf 130}, 813 (1963).
D.Heckathorn, Nucl.Phys. {\bf B156}, 328 (1979).
G.S.Adkins, Phys.Rev. {\bf D27}, 1814 (1983); {\bf D34}, 2489 (1986).
P.J.Doust, Ann.Phys.(N.Y.) {\bf 177}, 169 (1987).
P.J.Doust and J.C.Taylor, Phys.Lett. {B197}, 232 (1987).
J.C.Taylor, in ``Physical and Nonstandard Gauges", eds. P.Gaigg, W.Kummer and
M.Schweda, Lecture Notes Phys. n.361 (Springer, Berlin, 1990), p.137.
G.Leibbrandt, ``Non-Covariant Gauges", ch.9 (World Scientific, Singapore, 1994).
\bibitem{lav}
M.Lavelle and D.McMullan, Phys.Rep. {\bf C279},1 (1997).
E.Bagan, M.Lavelle, D.McMullan, B.Fiol and N.Roy, ``How do Constituent Quarks
arise in QCD? Perturbation Theory and the Infra-Red, talk at QCD-96, Montpellier
1996. R.Horan, M.Lavelle and D.McMulla, ``Charges in Gauge Theories", Univ.
Plymouth research report PLY-MS-48 (1998).
E.Bagan, M.Lavelle and D.McMullan, Phys.Lett. {\bf B370}, 128 (1996);
``A Class of Physically Motivated Gauges with
an Infra-Red Finite Electron Propagator", preprint UAB-FT-384, PLY-MS-96-01
(HEP-TH/9602083).


\end{references}
\end{document}